\documentclass[12pt]{article}
\pdfoutput=1

\usepackage[bulletsep]{collref}
\usepackage{amssymb,graphicx}
\usepackage[intlimits]{amsmath}
\usepackage{bbm}
\usepackage[small]{subfigure}

\usepackage{MnSymbol}

\makeatletter \@addtoreset{equation}{section} \makeatother

\makeatletter
\let\old@startsection=\@startsection
\let\oldl@section=\l@section
\renewcommand{\@startsection}[6]{\old@startsection{#1}{#2}{#3}{#4}{#5}{#6\mathversion{bold}}}
\renewcommand{\l@section}[2]{\oldl@section{\mathversion{bold}#1}{#2}}
\makeatother

\makeatletter
\let\old@makecaption=\@makecaption
\def\@makecaption{\small\old@makecaption}
\makeatother

\usepackage{color}

\renewcommand{\leq}{\leqslant}

\newcommand{\Q}{Q}

\begin{document}


\thispagestyle{empty}
\begin{flushright}\footnotesize
\texttt{NORDITA 2021-151} 
\vspace{0.6cm}
\end{flushright}

\renewcommand{\thefootnote}{\fnsymbol{footnote}}
\setcounter{footnote}{0}

\begin{center}
{\Large\textbf{\mathversion{bold} Integrable domain walls in ABJM theory}
\par}

\vspace{0.8cm}

\textrm{Charlotte~Kristjansen$^{1}$, Dinh-Long Vu$^{2}$ and
Konstantin~Zarembo$^{1,2}$\footnote{Also at ITEP, Moscow, Russia}}
\vspace{4mm}

\textit{${}^1$Niels Bohr Institute, Copenhagen University, Blegdamsvej 17, \\ 2100 Copenhagen, Denmark}\\
\textit{${}^2$Nordita, KTH Royal Institute of Technology and Stockholm University,
Hannes Alfv\'{e}ns V\"{a}g 12, 114 19 Stockholm, Sweden}\\
\vspace{0.2cm}
\texttt{kristjan@nbi.dk, dinh-long.vu@su.se, zarembo@nordita.org}

\vspace{3mm}


\par\vspace{1cm}

\textbf{Abstract} \vspace{3mm}

\begin{minipage}{13cm} 

One-point functions of local operators are studied,
at weak and strong coupling, for the ABJM theory in the presence of
a 1/2 BPS domain wall. In the underlying quantum spin chain the
domain wall is represented by a
boundary state which we show is integrable
yielding a compact determinant formula
for one-point functions of generic operators.

\end{minipage}
\end{center}

\vspace{0.5cm}

\newpage
\setcounter{page}{1}
\renewcommand{\thefootnote}{\arabic{footnote}}
\setcounter{footnote}{0}
\newcommand{\Cset}{{\,\,{{{^{_{\pmb{\mid}}}}\kern-.47em{\mathrm C}}}}}

\section{Introduction}
The study of domain wall set-ups featuring Nahm poles in ${\cal N}=4$ SYM has provided us with novel examples of   integrable boundary states
which, owing to the integrable 
$PSU(2,2|4)$ super spin chain underlying the $AdS_5/CFT_4$ 
correspondence~\cite{Beisert:2010jr},  have a discrete realization as matrix product states or valence bond states~\cite{deLeeuw:2015hxa,Buhl-Mortensen:2015gfd,Kristjansen:2020mhn}.  
Overlaps between these integrable boundary states 
and Bethe eigenstates encode the one-point functions of the field theory in the presence of the domain wall. For the simplest1/2 BPS D3-D5 domain wall the  overlaps are now known in a closed form at any loop order~\cite{Gombor:2020kgu,Gombor:2020auk,Komatsu:2020sup} and at least in a certain sub-sector non-perturbatively as well~\cite{Komatsu:2020sup}. Prior to
this, numerous partial results were obtained at fixed loop order and in particular sub-sectors and served  as input for the bootstrap idea leading to the progress above~\cite{deLeeuw:2015hxa,Buhl-Mortensen:2015gfd,Kristjansen:2020mhn,deLeeuw:2016umh,DeLeeuw:2018cal,Buhl-Mortensen:2017ind}. These results were based on the powerful machinery of boundary integrability in quantum spin chains, and recent progress in this area \cite{Gombor:2021hmj} will be instrumental in our analysis as well.
For a more involved non-supersymmetric D3-D7 set-up the overlaps are
known at the leading loop order~\cite{DeLeeuw:2019ohp}.

A fully non-perturbative analysis of the integrability of the boundary conditions would most likely require a treatment of the overlaps by means of the quantum spectral curve~\cite{Gromov:2014caa}
the basics of which is encoded in the QQ-system for the integrable super spin chain~\cite{Tsuboi:2009ud}. An analysis of the overlap formulas of the D3-D5 domain wall model from the point of view of the QQ-system
revealed that these formulas were singled out by exhibiting specific covariance properties under the fermionic duality transformations implied by the QQ-system~\cite{Kristjansen:2020vbe}.  This raised the hope that a covariance criterion could be used to constrain or even to fully determine yet unknown overlap formulas. 

Another system where one could imagine closed overlap formulas to exist is a domain wall version of ABJM theory. First of
all the $AdS_4/CFT_3$ system has a description as an integrable $OSp(6|4)$ super spin chain~\cite{Minahan:2008hf, Gromov:2008qe} and secondly integrable matrix
product states were already discovered in the study of determinant operators in ABJM theory~\cite{Yang:2021hrl}
as they were in ${\cal N}=4$ SYM theory~\cite{Jiang:2019zig}.

In the present paper we point out that there exists a 1/2 BPS domain wall version of ABJM theory, with a string theory dual
taking the form of a D2-D4 probe brane system with flux, which shares many characteristics with the D3-D5 domain wall version
of ${\cal N}=4$ SYM and for which the one-point functions can be found in closed form. The BPS conditions can
again be expressed as a set of Nahm equations and the one-point functions can be calculated both at strong and at weak coupling
(so far at the leading order). Furthermore, in both cases the probe D-brane configuration without flux corresponds to an integrable boundary condition according to an analysis carried out by Dekel and Oz~\cite{Dekel:2011ja}.
We determine the overlap formula of the ABJM domain wall model in the scalar sector by exploiting a newly derived result
for overlaps in a class of bosonic spin chains~\cite{Gombor:2021hmj} and use the requirement of covariance under fermionic dualities to uniquely fix the formula for the full ABJM theory.  We also perform computations of one-point functions from the 
string theory perspective.

Our paper is organized as follows. We begin by presenting the domain wall version of ABJM theory in
section~\ref{field-theory} and  in particular turning the BPS condition into  Nahm equations for certain 
field combinations. Then we turn to describing the dual  string theory set-up in section~\ref{string-conf}. The subsequent two sections are devoted to the calculation of one-point functions in the field theory language addressing protected states in section~\ref{weak} and excited states in section~\ref{excited}. Section~\ref{fluctuations} concerns the calculation of one-point functions in the string theory language. Finally, section~\ref{conclusion} contains our conclusion and outlook.

\section{The field theory set-up \label{field-theory}}

\subsection{A classical solution of the BPS equations}

The ABJM model \cite{Aharony:2008ug} is a Chern-Simons-matter theory in three dimensions with  $U(N)\times U(N)$ gauge symmetry and the Lagrangian \cite{Benna:2008zy}
\begin{eqnarray}\label{ABJM-L}
 \mathcal{L}&=&\frac{k}{4\pi }\,\mathop{\mathrm{tr}}\left[
 \varepsilon ^{\mu \nu \lambda }\left(
 A_\mu \partial _\nu A_\lambda +\frac{2}{3}\,A_\mu A_\nu A_\lambda 
 -\hat{A}_\mu \partial _\nu \hat{A}_\lambda -\frac{2}{3}\hat{A}_\mu \hat{A}_\nu \hat{A}_\lambda 
 \right)
 \right.
\nonumber \\
&&\left.\vphantom{\varepsilon ^{\mu \nu \lambda }\left(
 A_\mu \partial _\nu A_\lambda +\frac{2}{3}\,A_\mu A_\nu A_\lambda 
 -\hat{A}_\mu \partial _\nu \hat{A}_\lambda -\hat{A}_\mu \hat{A}_\nu \hat{A}_\lambda 
 \right)}
  +D_\mu Y^\dagger _AD^\mu Y^A
 +\frac{1}{12}\,Y^AY^\dagger _AY^BY^\dagger _BY^CY^\dagger _C
 +\frac{1}{12}\,Y^AY^\dagger _BY^BY^\dagger _CY^CY^\dagger _A
 \right.
\nonumber \\
&&\left.\vphantom{\varepsilon ^{\mu \nu \lambda }\left(
 A_\mu \partial _\nu A_\lambda +\frac{2}{3}\,A_\mu A_\nu A_\lambda 
 -\hat{A}_\mu \partial _\nu \hat{A}_\lambda -\frac{2}{3}\,\hat{A}_\mu \hat{A}_\nu \hat{A}_\lambda 
 \right)}
  -\frac{1}{2}\,Y^AY^\dagger _AY^BY^\dagger _CY^CY^\dagger _B
 +\frac{1}{3}\,Y^AY^\dagger _BY^CY^\dagger _AY^BY^\dagger _C
 +{\rm fermions}
 \right].
\end{eqnarray}
The scalars $Y^A$ belong to the bi-fundamental representation of the gauge group and transform as $\mathbf{4}$ of the $SU(4)$ R-symmetry: $A=1,\ldots, 4$. At large $N$ and fixed 't~Hooft coupling
\begin{equation}
 \lambda =\frac{N}{k},
\end{equation}
the model is dual to string theory on $AdS_4\times \mathbb{C}P^3$ with the dimensionless string tension $T\simeq \sqrt{\lambda /2}$  \cite{Aharony:2008ug}.

The domain wall (at weak coupling) is described by a classical scalar-field profile that satisfies the equations of motions. More precisely, it must fulfill a set of simpler BPS equations, in order to preserve part of the supersymmetry.  The energy of a BPS configuration takes the smallest possible value allowed by the boundary conditions. Once the energy density is represented as a total square plus a total derivative yielding a boundary term, the BPS condition requires the total square to vanish. The potential energy in (\ref{ABJM-L}) is not a total square in general, but becomes such when evaluated on a configuration with only two fields excited. We denote those $Y^1$ and $Y^2$, or collectively $Y^\alpha $. 

This follows from an identity
\begin{eqnarray}
 \mathop{\mathrm{tr}}Y^\dagger _\alpha Y^\beta Y^\dagger _\gamma Y^ \alpha Y^\dagger _\beta Y^\gamma 
 &=&3\mathop{\mathrm{tr}}Y^\dagger _\alpha Y^\beta Y^\dagger _\beta Y^\alpha Y^\dagger _\gamma Y^\gamma 
 -\mathop{\mathrm{tr}}Y^\dagger _\alpha Y^\beta Y^\dagger _\beta Y^\gamma Y^\dagger _\gamma Y^\alpha 
\nonumber \\
&&
 -\mathop{\mathrm{tr}}Y^\dagger _\alpha Y^\alpha Y^\dagger _\beta Y^\beta Y^\dagger _\gamma Y^\gamma ,
\end{eqnarray}
 derived by
contracting 
$\varepsilon ^{\alpha \beta \gamma }\varepsilon _{\delta \varepsilon \omega }=0$ with $Y^\dagger _\alpha Y^\delta Y^\dagger _\beta Y^\varepsilon Y^\dagger _\gamma Y^\omega $, and thus is only valid for $\alpha ,\beta ,\ldots =1,2$. One of the tensor structures in (\ref{ABJM-L}) becomes redundant and the horribly looking energy functional collapses into a neat modulus-squared form:  
\begin{eqnarray}\label{wall-energy}
 E&=&\frac{k}{4\pi }\int_{}^{}dx\,
 \mathop{\mathrm{tr}}
 \left(
 \frac{dY^\dagger _\alpha}{dx} \mp\frac{1}{2}Y^\dagger _\beta  Y^\beta Y^\dagger _\alpha  \pm\frac{1}{2}Y^\dagger _\alpha Y^\beta Y^\dagger _\beta
 \right)
\nonumber \\ &&\times 
 \left(
 \frac{dY^\alpha}{dx} 
 \mp\frac{1}{2}Y^\alpha  Y^\dagger _\beta Y^\beta 
 \pm\frac{1}{2}Y^\beta Y^\dagger _\beta Y^\alpha 
 \right)+{\rm total~derivative},
\end{eqnarray}
where $x\equiv x_2$ is the coordinate transverse to the domain wall. Either sign can be taken, corresponding to BPS and anti-BPS solutions. We take the upper sign for definiteness. 

The BPS equations correspond to the absolute minimum of the energy in a given "topological sector", when the integrand is set to zero altogether:
\begin{equation}\label{BPS}
 \frac{dY^\alpha}{dx} =
 \frac{1}{2}Y^\alpha  Y^\dagger _\beta Y^\beta 
 -\frac{1}{2}Y^\beta Y^\dagger _\beta Y^\alpha .
\end{equation}
 In \cite{Terashima:2008sy}
these equations  were obtained  by demanding the supersymmetry variation of the fermions to vanish. The solution will consequently host fermion zero modes and will automatically preserve half of the supersymmetry. 

The solution found in \cite{Terashima:2008sy} has a scale-invariant form
\begin{equation}\label{Terashima1}
 Y^\alpha =\frac{S^\alpha }{\sqrt{x}}\,, \hspace{0.5cm} x>0
\end{equation}
where $S^\alpha $ are two rectangular matrices   
\begin{equation}\label{Terashima2}
 S^1_{ij}=\delta _{i,j-1}\sqrt{i},\qquad S^2_{ij}=\delta _{ij}\sqrt{q-i}\,,\qquad i=1,\ldots ,q-1\qquad j=1,\ldots ,q.
\end{equation}
Assuming $q\leq N$ such a solution can  be placed in the upper 
left corner of   the $N\times N$ matrix fields $Y^A$ with $A=1,2$.  The resulting configuration can be checked to solve the BPS equations. It breaks the gauge symmetry down to $U(N-q+1)\times U(N-q)$ for finite values of $x$ whereas asymptotically
as $x\rightarrow +\infty $ the symmetry becomes $U(N)\times U(N)$.
 On the other side of the domain wall, $x<0$, the gauge symmetry is taken
to be $U(N-q+1)\times U(N-q)$.
 Incidentally, the same matrices $S^\alpha $ describe the  non-Abelian Coulomb branch of the mass-deformed ABJM theory\cite{Basu:2004ed}.

\subsection{Nahm's equations}

Before proceeding, we would like to make contact with Nahm's equations \cite{Nahm:1979yw} that describe supersymmetric domain walls in 4D \cite{Diaconescu:1996rk,Constable:1999ac} and naturally arise in $\mathcal{N}=4$ super-Yang-Mills theory with boundaries or defects \cite{Gaiotto:2008ak}. Nahm's equations have a clear geometric interpretation in the large-$N$ limit encapsulating the spherical shape of the brane embedded in $ S^5$ through the fuzzy, non-commutative geometry of the solution at large but finite $N$.
We will argue that the BPS equation (\ref{BPS}) is, in some sense, a square root of the Nahm's equation.

Consider, to this end, a composite field
\begin{equation}\label{Phi}
 \Phi ^\alpha _{\hphantom{\alpha }\beta }=Y^\alpha Y^\dagger _\beta .
\end{equation}
Assuming that $Y^\alpha $ satisfy the BPS condition  (\ref{BPS}), we  can differentiate $ \Phi ^\alpha _{\hphantom{\alpha }\beta }$ after $x$ to get a closed system of equations:
\begin{equation}
 \frac{d \Phi ^\alpha _{\hphantom{\alpha }\beta }}{dx}=
 \Phi ^\alpha _{\hphantom{\alpha }\gamma }
  \Phi ^\gamma  _{\hphantom{\alpha }\beta }
  -\frac{1}{2}\,
  \left\{ \Phi ^\gamma _{\hphantom{\alpha }\gamma }, 
  \Phi ^\alpha _{\hphantom{\alpha }\beta }\right\}.
\end{equation}
These equations can be further simplified by expanding the composite field in the $(\boldsymbol{\sigma },\mathbbm{1})$ basis:
\begin{equation}\label{sigma-exp}
  \Phi ^\alpha _{\hphantom{\alpha }\beta }=\Phi ^i\sigma ^\alpha _{i\,\beta }
  + \Phi \delta ^\alpha _{\hphantom{\alpha }\beta }.
\end{equation}
For the expansion coefficients we find, after simple algebra:
\begin{eqnarray}
\label{Nahm}
 \frac{d\Phi ^i}{dx}&=&\frac{i}{2}\,\varepsilon ^{ijk}[\Phi ^j,\Phi ^k],
\\
\label{auxi-Nahm}
 \frac{d\Phi }{dx}&=& \Phi ^i\Phi ^i-\Phi ^2.
\end{eqnarray}
The first equality is the Nahm equation.

The BPS domain wall (\ref{Terashima1}) corresponds to the simplest Nahm-pole solution:
\begin{equation}\label{Nahm-pole-Phi}
 \Phi ^i=\frac{t^i}{x}\,,
\end{equation}
where, in virtue of (\ref{Nahm}), $t^i$ must satisfy the $\mathfrak{su}(2)$ commutation relations:
\begin{equation}
 [t^i,t^j]=i\varepsilon ^{ijk}t^k.
\end{equation}
The $(q-1)\times (q-1)$ matrices $t^i$ thus form a $(q-1)$-dimensional representation of $\mathfrak{su}(2)$. The Casimir $\Phi ^i\Phi ^i$ determines the singlet component $\Phi $ through eq.~(\ref{auxi-Nahm}):
\begin{equation}
 \Phi =\frac{q\mathbbm{1}}{2x}\,.
\end{equation}

The connection to the $\mathfrak{su}(2)$ representation theory is quite fascinating. And it fits well with the dual supergravity description. The D4-brane representing the domain wall wraps a $\mathbbm{C}P^1=S^2$ in $\mathbbm{C}P^3$, and a large $\mathfrak{su}(2)$ representation can be interpreted as a fuzzy two-sphere that becomes smooth in the  large-$q$ limit. The
appearance of a fuzzy two-sphere in connection with the BPS equations of ABJM theory was also discussed in~\cite{Nastase:2009ny,Nastase:2009zu}. 

The same chain of arguments applies to the dual bi-linear:
\begin{equation}\label{dbilineal}
 \hat{\Phi }_\alpha ^{\hphantom{\alpha }\beta }
 =Y^\dagger _\alpha Y^\beta \equiv \hat{\Phi }_i\sigma ^{i\,\beta }_\alpha +\hat{\Phi }\delta _\alpha ^{\hphantom{\alpha }\beta }.
\end{equation}
The triplet component again satisfies Nahm's equations and the singlet is determined by the Casimir $\hat{\Phi} _i\hat{\Phi} _i$:
\begin{eqnarray}
\label{newNahm}
 \frac{d\hat{\Phi} _i}{dx}&=&-\frac{i}{2}\,\varepsilon _{ijk}[\hat{\Phi} _j,\hat{\Phi} _k],
\\
\label{new-auxi-Nahm}
 \frac{d\hat{\Phi} }{dx}&=& -\hat{\Phi} _i\hat{\Phi} _i+\hat{\Phi} ^2.
\end{eqnarray}
The Nahm-pole solution describing the domain wall is
\begin{eqnarray}
 \hat{\Phi }_i&=&-\frac{\hat{t}_i}{x}\,,\qquad \qquad [\hat{t}_i,\hat{t}_j]=i\varepsilon _{ijk}\hat{t}_k,
\nonumber \\
\hat{\Phi }&=&\frac{(q-1)\mathbbm{1}}{2x}\,.
\end{eqnarray}

The simplest domain wall solution has $q=2$. The bilinear (\ref{Phi}) then is a $1\times 1$ matrix (only its $11$ component is non-zero), and since in the 1d representation $t^i$ are trivial the bi-linear field takes a super-simple form:
\begin{equation}
 \Phi ^\alpha _{\hphantom{\alpha }\beta }=\frac{\delta^\alpha _{\hphantom{\alpha }\beta } }{x}\qquad (q=2).
\end{equation}
The dual bilinear (\ref{dbilineal}) is a $2\times 2$ matrix:
\begin{equation}
  \hat{\Phi }_\alpha ^{\hphantom{\alpha }\beta }=\frac{\mathbbm{1}\,\delta_\alpha ^{\hphantom{\alpha }\beta }-\sigma _3\sigma _i\sigma _3\,\sigma ^{i\,\beta }_\alpha  }{2x} \qquad (q=2).
\end{equation}

\section{The dual string configuration \label{string-conf}}

We expect the string theory configuration dual to the BPS solution of the previous section to be
a D2-D4 probe brane system in type IIA superstring theory where the probe brane has
geometry $AdS_3\times \mathbb{C}P^1\subset AdS_4\times \mathbb{C}P^3$
and carries $q$ units of world volume gauge field flux on the $\mathbb{C}P^1$. We can find such a probe brane embedding by extremizing the DBI plus WZ action for a D4-brane in the $AdS_4\times \mathbb{C}P^3$ background as outlined in~\cite{Chandrasekhar:2009ey}. The  probe brane embedding involving the wrapping of a $\mathbb{C}P^1 \subset 
 \mathbb{C}P^3$ matches the $SU(2)\times SU(2) \times U(1)$ symmetry of the vevs of the scalar fields of the Chern Simons field theory and, as we shall see, the additional flux
accounts for the jump in the rank of the gauge group across the domain wall.

\subsection{Choice of metric}
Let us start by choosing an appropriate parametrization of $S^7$. We follow reference~\cite{Nishioka:2008gz} and introduce four complex variables as follows
\begin{eqnarray}
Z^1=\cos\left({\xi}\right)\cos\left(\frac{\theta_1}{2}\right) e^{i(\chi_1+\phi_1)/2},&
&Z^2=\cos\left({\xi}\right)\sin\left(\frac{\theta_1}{2}\right) e^{i(\chi_1-\phi_1)/2}, \nonumber\\
Z^3=\sin\left({\xi}\right)\cos\left(\frac{\theta_2}{2}\right) e^{i(\chi_2+\phi_2)/2},&
&Z^4=\sin\left({\xi}\right)\sin\left(\frac{\theta_2}{2}\right)e^{i(\chi_2-\phi_2)/2},\nonumber
\end{eqnarray}
where $\xi\in[0,\frac{\pi}{2}[$ and $\theta_1,\theta_2\in [0,\pi]$. Furthermore, $\phi_1,\phi_2\in [0,2\pi]$ and $\chi_1,\chi_2\in [0,4\pi[$.
With this parametrization the metric of $S^7$ can be written as
\begin{eqnarray}
ds^2_{S^7}&=& d\xi^2+\frac{\cos^2\xi}{4}\left[(d\chi_1+\cos\theta_1 d\phi_1)^2+d \theta_1^2+\sin^2\theta_1 d\phi_1^2\right]
\\
&&+\frac{\sin^2\xi}{4}\left[(d\chi_2+\cos\theta_2 d\phi_2)^2+d \theta_2^2+\sin^2\theta_2\, d\phi_2^2\right]. \nonumber
\end{eqnarray}
Next, we define new coordinates by
\begin{equation}
\chi_1=2y+\psi, \hspace{0.5cm} \chi_2=2y-\psi,
\end{equation}
where $y\in [0,2\pi]$, $\psi\in [-2\pi,2\pi]$. Then we can implement the quotient $S^7/\mathbb{Z}_k$ by making the identification
\begin{equation}
y\sim y+ \frac{2\pi}{k}.
\end{equation}
We can now also rewrite the metric of $S^7$ as
\begin{equation}
ds^2_{S^7}= ds_{\mathbb{C}P^3}^2+(dy+A)^2, 
\end{equation}
where
\begin{equation}
A=\frac{1}{2}(\cos^2 \xi-\sin^2\xi) d\psi+\frac{1}{2}\cos^2\xi \cos\theta_1 d\phi_1+\frac{1}{2}\sin^2\xi \cos \theta_2 d\phi_2,
\end{equation}
and
\begin{eqnarray}
ds_{\mathbb{C}P^3}^2&=&d\xi^2 +\cos^2\xi \sin^2\xi \left(d\psi+\frac{\cos\theta_1}{2} d\phi_1-\frac{\cos \theta_2}{2} d\phi_2\right)^2 \\
&&+\frac{1}{4}\cos^2 \xi \left(d\theta_1^2+\sin^2\theta_1 d\phi_1^2\right)
+\frac{1}{4}\sin^2 \xi \left(d\theta_2^2+\sin^2\theta_2 d\phi_2^2\right). \nonumber
\end{eqnarray}
For $AdS_4$ we can use the Poincar\'{e} metric 
\begin{equation}
ds^2_{AdS_4}= \frac{1}{z^2}dz^2+z^2(-dx_0^2+dx_1^2+dx_2^2),
\end{equation}
where the AdS boundary is at $z\rightarrow \infty$.

 The relevant 10-dimensional background of type IIA string theory
is described by the metric
\begin{equation}
ds^2={\widetilde{R}^2}(ds^2_{AdS_4}+ 4 ds^2_{\mathbb{C}P^3}),
\end{equation}
where 
\begin{equation}
\frac{\widetilde{R}^2}{\alpha'}=\pi \sqrt{\frac{2N}{k}}. \label{TildeR}
\end{equation}
The background comes with a RR 2-form field as well as a RR 4-form field which are given by
\begin{eqnarray}
F^{(2)}= k\left(-\cos \xi \sin \xi d\xi \wedge \left( 2 d\psi +\cos \theta_1 d\phi_1-\cos \theta_2 d\phi_2\right) \right.\nonumber\\
\left.
-\frac{1}{2} \cos^2\xi \sin \theta_1 d\theta_1 \wedge d\phi_1-\frac{1}{2} \sin^2\xi \sin \theta_2 d\theta_2 \wedge d\phi_2\right),
\end{eqnarray}
\begin{equation}
F^{(4)}=\frac{3 R^3}{8} \epsilon_{AdS_4},
\end{equation}
where $\epsilon_{AdS_4}$ is the volume form on $AdS_4$, and $\left(\frac{R}{l_p}\right)^3=4k\left(\frac{\widetilde{R}}{\sqrt{\alpha'}}\right)^2$.
\subsection{The probe brane embedding}
We are interested in a D4 probe with geometry $AdS_3\times  \mathbb{C}{P}^1$ embedded in the IIA background. 
We make the ansatz
that the probe brane is placed at $\xi=0$ (and $\theta_2,\phi_2,\psi$ constant) and we take its world volume coordinates
to be $z, x_0,x_1, \theta_1, \phi_1$ while the last embedding  coordinate $x_2$ is supposed to be non-constant but
to depend only on $z$. The $\xi=0$ condition singles out a $\mathbb{C}P^1\subset \mathbb{C}P^3$ wrapped by the brane and parametrized by the coordinates $\theta_1,\phi_1$. The brane likewise wraps an $AdS_3\subset AdS_4$ parametrized by the coordinates
$z,x_0,x_1$. Furthermore, we turn on a world-volume gauge field on the $\mathbb{C}P^1$ given by the 2-form
\begin{equation}
{\cal F}=\widetilde{R}^2 Q \sin\theta_1 d\theta_1 \wedge d\phi_1.
\end{equation}
As we shall see, the parameter $Q$ is related to the rank of the representation of the classical fields in the
gauge theory, $q$. 
The probe Dp-brane action in general takes the form
\begin{eqnarray}
I&=&I_{DBI}+I_{WZ} \\
&=&-{T_p}\int d^{p+1}\sigma\,e^{-\Phi} \sqrt{-\det (G+{\cal F})}+T_p \int d^{p+1}\sigma \,e^{\cal F} \wedge \sum_m {\cal P}[C_m], \nonumber 
\end{eqnarray}
where $G$ is the induced metric on the brane, ${\cal F}$ is the world volume gauge field and the $C_m$'s are the various
RR background gauge field potentials where we have set the Kalb-Ramond field to zero. 
Furthermore ${\cal P}$ stands for the pull-back.
In the present case the $C_m$'s are $C_1$ and $C_3$, related to the field strengths via $F^{(4)}=dC_3$ and $F^{(2)}=dC_1$. In
our case only the term with $C_3$ will be non-vanishing and thus the Wess-Zumino terms reads
\begin{equation}
I_{WZ}=T_4\int d^5 \sigma \,{\cal F}\wedge {\cal P}[C_3].
\end{equation}
 Finding the induced metric $G$ is straightforward and we get
\begin{equation}
-\det (G+{\cal F})=\widetilde{R}^{10}\,z^2 \left(1+z^4\, (x_2'(z))^2\right)  \sin^2 \theta_1 \,(1+Q^2).
\label{inducedmetric}
\end{equation}
For $C_3$ we can take
\begin{equation}
C_3= \frac{R^3}{8}\,z^3 \, dx^0\wedge dx^1 \wedge dx^2,
\end{equation}
and we find
\begin{equation}
I_{WZ}= T_4\widetilde{R}^4 \, \frac{k}{2} \int d^5 \sigma \left( Q \sin \theta_1\, z^3 \,x_2'(z)\right).
\end{equation}
We now get for the total action
\begin{eqnarray}
I_{DBI+WZ}&=&T_4\, \frac{k}{2}\, \widetilde{R}^4\, \int_{-\infty}^{\infty}dx^0 \int_{-\infty}^\infty dx^1 \int_0^\pi d\theta_1 \sin \theta_1 \int_0^{2\pi} d\phi_1 \int_0^\infty dz\, \,{\cal  I} \nonumber
\\&=& {2\pi \sqrt{\alpha'} \,k}\,T_4 \, \widetilde{R}^4\, V\int_0^\infty dz \,{\cal I}, 
\end{eqnarray}
where $V=\int dx^0dx^1$ and
\begin{eqnarray}
{\cal I}&=&
\left[ -z \sqrt{(1+z^4(x_2'(z))^2)(1+Q^2)}+Q z^3 x_2'(z)\right], \nonumber
\end{eqnarray}
and where we have made use of the relation
 \begin{equation}
 e^{-\Phi}=\frac{k\sqrt{\alpha'}}{2\widetilde{R}}.
 \end{equation}
From this we get the following Euler Lagrange equation for $x_2(z)$
\begin{equation}
\frac{\partial}{\partial z} \left\{
\frac{z \sqrt{1+Q^2}}{\sqrt{1+z^4\, (x_2'(z))^2}} z^4 x_2'(z)-Qz^3\right\} =0
,
\end{equation}
and we see that it has the  solution
\begin{equation}
x_2(z)=\frac{Q}{z},
\end{equation}
which is the solution we expect to be relevant for our analysis. The parameter $Q$  should be associated with the number, $q$, of D2 branes ending on the D4 brane and thus the jump in
the rank of the gauge group across the defect in the field theory language. The relation can be found by interpreting the Wess-Zumino term of the D4 brane as the coupling of a number, $q$, of D2 branes  to the background gauge field $C^3$ and leads to
\begin{equation}
q=\frac{T_4}{T_2} \int_{S^2} {\cal F} =\frac{T_4}{T_2} \, 4\pi\widetilde{R}^2 Q =\frac{\widetilde{R}^2}{\pi \alpha'} Q=
\sqrt{\frac{2 N}{k}} Q,
\end{equation}
where we have made use of the relation
\begin{equation}
T_p=(2\pi)^{-p} (\alpha')^{-(p+1)/2},
\end{equation}
as well as eqn.~(\ref{TildeR}).
In particular we see that only if we take the parameter $Q$ finite will the probe brane have an angle with the $AdS_4$ boundary
which is different from $\frac{\pi}{2}$.  The probe brane configuration thus suggests the following double scaling limit 
\begin{equation}
\lambda \rightarrow \infty, \hspace{0.5cm} q\rightarrow \infty, \hspace{0.5cm}  \frac{2\lambda}{q^2}=Q^{-2} \label{dsl}
\hspace{0.5cm} \mbox{fixed}.
\end{equation}
This situation is reminiscent of the situation in the $AdS_5/CFT_4$ correspondence where similar Karch-Randall probe brane
set-ups of  D3-D5 and D3-D7 type likewise suggested the introduction of a double scaling limit involving the 't Hooft coupling constant in combination with a certain representation label~\cite{Nagasaki:2012re,Kristjansen:2012tn}. In the $AdS_5/CFT_4$
case this double scaling parameter allowed for a successful comparison of one-point functions of chiral primaries computed in gauge theory and in string theory to two leading orders both for a supersymmetric D3-D5 probe brane set-up~\cite{Nagasaki:2012re,Buhl-Mortensen:2016pxs,Buhl-Mortensen:2016jqo} and two non-supersymmetric D3-D7 probe brane set-ups~\cite{Kristjansen:2012tn,GimenezGrau:2018jyp,Gimenez-Grau:2019fld}. In the present case we shall see that a successful comparison is
only possible if an additional large charge limit is imposed. 

\subsection{Chiral operators at strong coupling \label{harmonics}}
One-point functions of chiral primary operators in the presence of a Karch-Randall probe brane can be calculated through a fluctuation analysis of the supergravity background.  The analysis will be carried out in section~\ref{fluctuations}.  Only chiral primaries whose symmetries are
compatible with the $SU(2)\times SU(2) \times U(1)$ symmetry of the brane embedding will have non-vanishing one-point functions. These particular chiral primaries depend only on the angular variable $\xi$ and are solutions  of  the Laplace equation
\begin{equation}
\nabla^2 Y= \frac{1}{\sqrt{g}}\, \partial_i\sqrt{g} \, g^{ij} \partial_j Y=-E Y, \label{Laplace}
\end{equation}
which reduces to
\begin{equation}
\nabla^2 Y= \frac{1}{\sin^3\xi \cos^3\xi} \,\partial_\xi \, \sin^3\xi \cos^3\xi \, \partial_\xi Y =-EY. \label{diffeqn}
\end{equation}
The most general solution of this equation  is a linear combination of a hypergeometric function
and a Meijers G-function. In order for the solution to be non-singular at $\cos \xi=0$ one has to discard the Meijers G-function,
and in order for the solution to be regular at $\cos \xi=1$ one needs
\begin{equation}
E=2\Delta (2\Delta+6),
\end{equation} 
where we recognize $\Delta$ as the conformal dimension of the corresponding operator. The solution more precisely reads
\begin{equation}
Y_{\Delta}(\xi)= {\cal N}_{\Delta} \cdot \, _2F_1\left[ -\Delta, \Delta+3,2, \cos^2\xi\right], \label{2F1}
\end{equation}
where ${\cal N}_{\Delta}$ is a normalization factor.

\section{One-point functions at weak coupling \label{weak}}

In analogy with the situation in ${\cal N}=4$ SYM~\cite{deLeeuw:2015hxa,Buhl-Mortensen:2015gfd},
tree level  one-point functions in the scalar sector of ABJM theory can be expressed as overlaps between Bethe eigenstates 
and Matrix Product States. Furthermore, as the bond dimension of the Matrix Product State becomes equal to one, the Matrix
Product State becomes a valence bond state. In the present section we expound these ideas and compute the one-point functions of protected scalar operators. The more intricate case of excited states is treated in the subsequent section.

\subsection{Boundary state overlaps}

The single-trace scalar operators of ABJM can be viewed as states in an alternating $SU(4)$ spin chain:
\begin{equation}\label{spin-chain}
 \mathcal{O}=\Psi _{A_1\,\ldots\, A_{2L-1}}^{\hphantom{A}A_2\,\ldots\, A_{2L}}\mathop{\mathrm{tr}}Y^{A_1}Y^\dagger _{A_2}\ldots Y^{A_{2L-1}}Y^\dagger _{A_{2L}}.
\end{equation}
The odd and even sites are occupied intermittently by $\mathbf{4}$ and $\bar{\mathbf{4}}$ of $SU(4)$.
The two-loop mixing matrix of scalar operators is identified with the spin-chain Hamiltonian \cite{Minahan:2008hf}:
\begin{equation}\label{1-loop-mixing}
 H=\lambda ^2\sum_{l=1}^{2L}\left(1-P_{l,l+2}+\frac{1}{2}\,P_{l,l+2}K_{l,l+1}+\frac{1}{2}\,K_{l,l+1}P_{l,l+2}\right),
\end{equation}
where $K_{lm}$ and $P_{lm}$ are the standard trace and permutation operators acting on sites $l$ and $m$:
\begin{equation}
 P^{A'B'}_{A\,\,B}=\delta ^{A'}_{B}\delta ^{B'}_A,\qquad 
 K^{A'\,A}_{B'\,B}=\delta ^{A'}_{B'}\,\delta ^{A}_{B}.
\end{equation}
The permutation always acts on the same type of spins, while the trace mixes the two  representations. 

Traceless and symmetric tensors (symmetric in each set of indices) define chiral primary operators. At fixed length they belong to the same representation of $\mathfrak{su}(4)$, the one with Dynkin labels $[L,0,L]$. All chiral primaries have zero energy and form the ground state multiplet of the spin-chain Hamiltonian. One can declare the lowest-weight state
\begin{equation}
 \mathcal{O}_{\rm vac}=\mathop{\mathrm{tr}}(Y^1Y^\dagger _2)^L,
\end{equation}
 the "true" vacuum and generate all other members of the multiplet by inserting zero-momentum, zero-energy excitations.
 
The domain wall induces non-zero one-point functions, which to the leading order in perturbation theory can be calculated by simply substituting the classical solution (\ref{Terashima1}), (\ref{Terashima2}) for the scalar fields $Y^A$ in (\ref{spin-chain}). The outcome can be expressed as an overlap of the operator's wavefunction with a fixed state  in the spin chain's Hilbert space, which we call MPS, the Matrix Product State:
\begin{equation}\label{O(x)}
 \left\langle \mathcal{O}(x)\right\rangle=\frac{1}{x^L}\,\,\frac{1}{\lambda ^LL^{\frac{1}{2}}}\,\,\frac{\left\langle {\rm MPS}\right.\!\left| \Psi \right\rangle}{\left\langle \Psi \right.\!\left|\Psi  \right\rangle^{\frac{1}{2}}}.
\end{equation}
The prefactor accounts for the difference between the spin-chain scalar product and the norm defined by the two-point correlation function.
The Matrix Product State is built from the matrices (\ref{Terashima2}) defining the classical solution:
\begin{equation}\label{gen-MPS}
 {\rm MPS} _{\hphantom{A}A_2\,\ldots\, A_{2L}}^{A_1\,\ldots\, A_{2L-1}}=\mathop{\mathrm{tr}}S^{A_1}S^\dagger _{A_2}\ldots S^{A_{2L-1}}S^\dagger _{A_{2L}}.
\end{equation}
The state so defined has non-zero components only when the indices $A_l$ take values $1$ or $2$ denoted collectively by greek letters as above.

Two convenient representations of the boundary state arise upon combining even and odd sites of the spin chain as in (\ref{Phi}) or (\ref{dbilineal}):
\begin{eqnarray}
 {\rm MPS}&=&\mathop{\mathrm{tr}}\nolimits_{{\rm aux}}M_{12}\ldots M_{2L-1,2L},
\nonumber \\
{\rm MPS}&=&\mathop{\mathrm{tr}}\nolimits_{{\rm aux}}\widehat{M}_{23}\ldots \widehat{M}_{2L,1}.
\end{eqnarray}
The building blocks are matrices in the auxiliary space tensored with a quantum state of two neighboring spins:
\begin{eqnarray}
 M&=&\frac{q}{2}\,\mathbbm{1}\otimes\mathbbm{1}+t_i\otimes\sigma ^i,
\nonumber \\
\widehat{M}&=&\frac{q-1}{2}\,\mathbbm{1}\otimes\mathbbm{1}-\hat{t}^i\otimes\sigma _i.
\end{eqnarray}
Here $t_i$ (or $\hat{t}^i$) and the first $\mathbbm{1}$ act in the auxiliary space of dimension $q-1$ (or $q$), while $\sigma ^i$ and the second $\mathbbm{1}$ represent the state in the quantum space of two neighboring sites of the spin chain. 
 
When $q=2$, one of the two $\mathfrak{su}(2)$ representations is trivial: $t_i=0$. There is no auxiliary space to trace over and the boundary state becomes two-site entangled:
\begin{equation}\label{MPS2}
 {\rm MPS}^{\hphantom{2\,\,}\alpha _1\,\ldots\, \alpha _{2L-1}}_{2\,\,\hphantom{\alpha }\alpha _2\,\ldots\, \alpha _{2L}}=
 \delta _{\hphantom{\alpha }\alpha _2}  ^{\alpha_1  }\,\ldots \,
 \delta ^{\alpha _{2L-1}}  _{\hphantom{\alpha }\alpha_{2L}  },
\end{equation}
in other words it is a Valence Bond State:
\begin{equation}\label{MPS->VBS}
 \left\langle {\rm MPS}_2\right|=\left\langle K\right|^{\otimes L}
 \equiv \left\langle {\rm VBS}\right|,
\end{equation}
where  $K$ is a two-site state with components
\begin{equation}\label{2-block}
 K^\alpha   _{\hphantom{\alpha }\beta  }=\delta _{\hphantom{\alpha }\beta}  ^{\alpha  }.
\end{equation}
Degeneration of a matrix product state into a valence bond state is reminiscent to a similar phenomenon in the $\mathfrak{su}(2)$ sector of $\mathcal{N}=4$ SYM \cite{Piroli:2017sei}.

\subsection{Selection rules}\label{selection}

We can now determine what type of single trace operators get non-vanishing one-point functions at tree-level. Obviously, the relevant operators must be built entirely from fields of the type $Y^1, Y^2, Y_1^\dagger, Y_2^{\dagger}$.  Among these is the chiral primary
\begin{equation}
{\cal O}=\mbox{Tr}(Y^1Y_2^\dagger \ldots Y^1Y_2^\dagger), \label{vacuum}
\end{equation}
the vacuum of the spin chain.
We find from (\ref{Phi}), (\ref{sigma-exp}) and (\ref{Nahm-pole-Phi}):
\begin{equation}
Y^1 Y_2^\dagger=\Phi ^1 _{\hphantom{1}2}=\frac{t^i\sigma ^1 _{i\,2}}{x}=\frac{2t^-}{x}\,,
\end{equation}
meaning that $Y^1 Y_2^\dagger$ is lower triangular and the traces of all of its powers vanish.
This is in contrast to the parallel study in ${\cal N}=4$ SYM where the corresponding chiral primary (vacuum of the spin chain) has a non-vanishing
one-point function~\cite{deLeeuw:2015hxa}. This fact should  be visible in a string theory analysis as well.
For the other two-site combinations of fields we find
\begin{eqnarray}
Y^2 Y_1^\dagger &=& \frac{2t^+}{x}, \\
Y^1 Y_1^\dagger &=&  \frac{q+2t^3}{2x},\\
Y^2 Y_2^\dagger &=&  \frac{q-2t^3}{2x}\,.
\end{eqnarray}
We thus see that operators with non-vanishing tree-level vevs can contain an arbitrary number of field combinations of the
the type $(Y^1 Y_1^\dagger)$ and $(Y^2 Y_2^\dagger)$ whereas a combination $(Y^1 Y_2^\dagger)$ requires 
another term of the type $(Y^2 Y_1^\dagger)$.

With the vacuum given by~(\ref{vacuum})
we see that in order to get a non-vanishing one-point function we need to have a number of excitations which is equal to half the length
of the spin chain. Among these, there is a particular symmetric set-up where we have the same number of excitations on the
odd sites of the spin chain as on the even sites.  Denoting the length of the alternating spin chain as 2L and using the 
distinguished Dynkin diagram of the underlying super Lie algebra these states are characterized by  having the number of Bethe roots (see below) on the three relevant nodes of the diagram identical and all equal to $L$:
\begin{equation}\label{Ki=L}
K_1=K_2=K_3=L.
\end{equation}
In order to evaluate the one-point functions of these operators we need to construct the appropriate eigenstates of the alternating spin chain. This problem was analyzed in~\cite{Minahan:2008hf}, see also~\cite{Yang:2021hrl}.

A state with $K_i$ Bethe roots belongs to the $\mathfrak{su}(4)$ representation with the Dynkin labels
\begin{equation}
[L-2K_1+K_2,K_1-2K_2+K_3,L-2K_3+K_2].
\end{equation}
Consequently, (\ref{Ki=L}) describes a singlet. This should not come as a surprise. The domain wall (equivalently, the boundary state in the spin chain) breaks $SU(4)$ to $SU(2)\times SU(2)\times U(1)$. An $SU(4)$ singlet is a singlet of this smaller group and is thus allowed to have a non-zero overlap with the boundary state. Singlets of $SU(2)\times SU(2)\times U(1)$  do exist in non-trivial representations of $SU(4)$ and they have non-zero one-point functions, but these states are not $SU(4)$ primaries and if we concentrate on highest weights, as typically done in the Bethe ansatz, then (\ref{Ki=L}) is a necessary condition for the one-point function  not to vanish.

An example of non-singlet operator with a non-trivial one-point function is a chiral primary
\begin{equation}
 \mathcal{O}_{\mathbf{n},\mathbf{m}}=\mathop{\mathrm{tr}}\left(n^\dagger _AY^A\,m^BY^\dagger _B\right)^L.
\end{equation}
The complex vectors $\mathbf{n}$ and $\mathbf{m}$ must satisfy $(\mathbf{n}^\dagger \cdot \mathbf{m})=0$. As follows from (\ref{O(x)}), (\ref{MPS2}),  for $q=2$ the one-point function is
\begin{equation}
 \left\langle \mathcal{O}_{\mathbf{n},\mathbf{m}}\right\rangle=\frac{1}{x^L}\,\,
 \frac{1}{\lambda ^LL^{\frac{1}{2}}}\,\,\frac{\left(\widetilde{\mathbf{n}}^\dagger \cdot \widetilde{\mathbf{m}}\right)^L}{\left(\mathbf{n}^\dagger \cdot \mathbf{n}\right)^{\frac{L}{2}}\left(\mathbf{m}^\dagger \cdot \mathbf{m}\right)^{\frac{L}{2}}}\,,
\end{equation}
where $\widetilde{\mathbf{n}}^\dagger =(n^\dagger _1,n^\dagger _2)$, $\widetilde{\mathbf{m}}=(m_1,m_2)$, i.e. complex
vectors with only two components.
Obviously $(\mathbf{n}^\dagger \cdot \mathbf{m})=0$ does not imply $\left(\widetilde{\mathbf{n}}^\dagger \cdot \widetilde{\mathbf{m}}\right)=0$. We now characterize $SU(2)\times SU(2) \times U(1)$-invariant chiral primaries and their one-point functions more precisely.

\subsection{Chiral primary operators at weak coupling}

The chiral primary operators are in one-to-one correspondence with the spherical harmonics on $\mathbb{C}P^3$ and only
those which carry the $SU(2)\times SU(2) \times U(1)$ symmetry of the classical fields will have non-vanishing one-point functions. As noticed earlier, there is only one chiral primary with this symmetry for each even value of the length. We can read off the relevant operators from the spherical harmonics that we determined in section~\ref{harmonics}. To do so we first rewrite these as homogeneous polynomials in $\cos\xi$ and $\sin \xi$. 
For the first few cases we get
\begin{eqnarray}
Y_1(\xi)&=&{\cal N}_1\left(-\cos^2\xi+\sin^2 \xi\right), \label{Ys} \\
Y_2(\xi)&=& {\cal N}_2\left(\cos^4 \xi-3\cos^2\xi \sin^2\xi +\sin^4 \xi\right), \nonumber \\
Y_3(\xi)&=&{\cal N}_3\left(-\cos^6\xi +6 \cos^4 \xi\sin^2 \xi- 6\sin^4 \xi \cos^2 \xi +\sin^6\xi\right), \nonumber \\
Y_4(\xi)&=&{\cal N}_4\left( \cos^8 \xi-10\cos^6 \xi\sin^2\xi+20 \cos^4 \xi\sin^4\xi -10 \cos^2 \xi \sin^6\xi+ \sin^8 \xi\right) . \nonumber
\end{eqnarray}
Secondly, we make the replacements $\cos^2\xi\rightarrow Y^1 {Y}_1^{\dagger}+Y^2Y_2^{\dagger}$ and
$\sin^2\xi\rightarrow Y^3 {Y}_3^{\dagger}+Y^4Y_4^{\dagger}$ and symmetrize the resulting powers of fields. 
For $Y_1(\xi)$ we trivially get
\begin{equation}
Y_1(\xi)={\cal N}_1\,\mbox{Tr}( Y^3 Y_3^\dagger +Y^4Y_4^\dagger-Y^1 Y_1^\dagger -Y^2Y_2^\dagger).
\end{equation}
Moving on to  $Y_2(\xi)$ we find
\begin{equation}
Y_2(\xi)={\cal N}_2\,C_{I_1 I_2}^{J_1 J_2}\, \,\mbox{Tr}\,Y^{I_1} {Y}_{J_1}^{\dagger}Y^{I_2} {Y}_{J_2}^{\dagger},
\end{equation}
with
\begin{eqnarray}
C_{1 1}^{1 1}=C_{22}^{22}=C_{33}^{33}=C_{44}^{44}&=&1, \nonumber \\
C_{1 2}^{ 12}=C_{12}^{21}=C_{21}^{12}=C_{21}^{21}=C_{34}^{ 34}=C_{34}^{43}=C_{43}^{34}=C_{43}^{43}&=&\frac{1}{2},
\nonumber \\
C_{1 3}^{ 13}=C_{13}^{31}=C_{31}^{13}=C_{31}^{31}=C_{14}^{ 14}=C_{14}^{41}=C_{41}^{14}=C_{41}^{41}&=&-\frac{3}{4}, 
\nonumber \\
C_{2 3}^{ 23}=C_{23}^{32}=C_{32}^{23}=C_{32}^{32}=C_{24}^{ 24}=C_{24}^{42}=C_{42}^{24}=C_{42}^{42}&=&-\frac{3}{4}
\nonumber
\end{eqnarray}
and thus $C$  being a symmetric traceless tensor as required. Only the components with indices taking values  1 or 2 will be of importance
for the one-point functions but all components play a role when it comes to normalization. Requiring the operators to be
unit normalized, i.e.
\begin{equation}
C_{I_1 I_2\ldots I_{L}}^{J_1 J_2\ldots J_{L}} \,C^{I_1 I_2\ldots I_{L}}_{J_1 J_2\ldots J_{L}}=1
\end{equation}
gives the following normalization constant
\begin{equation}
{\cal N}_{L}= \left( \begin{matrix} 2L+2 \\ L \end{matrix}\right)^{-1/2},
\end{equation}
and leads to the following  normalization of the spherical harmonics on the string theory side
\begin{align}
\int_{S^7/\mathbb{Z}_k} Y_L(Y_K)^*=\delta_{LK}2\pi^4\frac{(L!)^2}{(2L+3)!}.
\end{align}
The one point function of these chiral primaries can be found by direct computation and reads for the
first few cases (excluding the field theoretical prefactor that is reinstated below in eqn.~(\ref{general}))
\begin{align}
\langle Y_1\rangle&=\frac{1}{2}(q^2-q), \\
\langle Y_2\rangle&=\frac{1}{\sqrt{15}}(q^3-\frac{3}{2}q^2+\frac{1}{2}q), 
\label{length-2} \\
\langle Y_3\rangle&=\frac{1}{\sqrt{56}}(q^4-2q^3+q^2), \\
\langle Y_4\rangle&=\frac{1}{\sqrt{210}}(q^5-\frac{5}{2}q^4+\frac{5}{3}q^3-\frac{1}{6}q) \\
\langle Y_5\rangle&=\frac{1}{\sqrt{792}}(q^6-3q^5+\frac{5}{2}q^4-\frac{1}{2}q^2), \\
\langle Y_6\rangle &=\frac{1}{\sqrt{3003}}(q^7-\frac{7}{2}q^6+\frac{7}{2}q^5-\frac{7}{6}q^3+\frac{1}{6}q), 
\end{align}
which is compatible with the following general expression.
\begin{align}
\langle Y_L\rangle=\frac{1}{x^L}\,\frac{1}{\lambda^L L^{1/2}}\,\mathcal{N}_L(L+1)\sum_{a=1}^{q-1}a^L. \label{general}
\end{align}


\section{Excited states \label{excited}}

We now turn to studying one-point functions of non-protected operators starting with the scalar sector in 
subsection~\ref{SU(4)} and moving 
on the full theory in subsection~\ref{Full}.

\subsection{$SU(4)$ sector\label{SU(4)}}

The spectrum of the $SU(4)$ spin chain (\ref{1-loop-mixing}) is described by the  
standard group-theoretic Bethe equations 
\begin{eqnarray}
&&\left(\frac{u_{aj}-\frac{iq_a}{2}}{u_{aj}+\frac{iq_a}{2}}\right)^L
 \prod_{bk}^{}\frac{u_{aj}-u_{bk}+\frac{iM_{ab}}{2}}{u_{aj}-u_{bk}-\frac{iM_{ab}}{2}}
 \equiv \,{\rm e}\,^{i\chi _{aj}}=\left(-1\right)^{\frac{M_{aa}}{2}},
\\
&&
E=\lambda ^2\sum_{aj}^{}\frac{q_{a}}{u_{aj}^2+\frac{q_a^2}{4}}\,,
\end{eqnarray}
specified to the $SU(4)$ Cartan matrix and the Dynkin labels of the spin representation:
\begin{equation}
 M=\begin{bmatrix}
 2  & -1 &  0 \\ 
 -1  & 2 & -1  \\ 
  0  & -1 &  2 \\ 
 \end{bmatrix},\qquad 
 q=\begin{bmatrix}
  1 \\ 
  0 \\ 
  1 \\ 
 \end{bmatrix}.
\end{equation}

The hallmark of boundary integrability is a $\mathbbm{Z}_2$ symmetry acting on the Bethe roots
\begin{equation}\label{parity}
 \Omega :~u_{aj}\rightarrow -u_{\sigma (a)j}.
\end{equation}
Only invariant Bethe states have a non-zero overlap with the boundary state at hand, (\ref{gen-MPS}) in our case. This can be regarded a defining property of a boundary state that makes it consistent with the underlying integrability structure \cite{Ghoshal:1993tm,Piroli:2017sei}.

The boundary state is called chiral if parity acts trivially on the Dynkin diagram ($\sigma = {\rm id}$), and achiral otherwise \cite{MacKay:2011zs,Gombor:2020kgu}. The permutation $\sigma $ should anyway square to the identity: $\sigma ^2={\rm id}$ and be a symmetry of the Bethe equations: $M_{\sigma (a)\sigma (b)}=M_{ab}$ and $q_{\sigma (a)}=q_a$. The boundary state we consider is actually achiral, with parity interchanging the momentum-carrying nodes:
\begin{equation}\label{(13)reflection}
 \sigma =(13).
\end{equation}
The $\mathbbm{Z}_2$ transformation represents the spacial reflection of the spin chain, or charge conjugation in terms of the original field-theory operators. The ABJM spin chain  hosts complex representations $\mathbf{4}$ and $\bar{\mathbf{4}}$ at alternate sites and those get interchanged by parity. The reflection (\ref{(13)reflection}) is the complex conjugation acting on the Dynkin diagram and therefore must accompany inversion of rapidities.
 
Parity ordains a symmetrically paired structure of the Bethe roots:\\ $\{u_{aj},-u_{\sigma (a)j}\}$. Solitary zero roots are also allowed on the nodes 
with $\sigma (a)=a$, being fixed points of parity $\Omega $. The key word here is solitary. Paired roots can accidentally 
take zero values, namely it can happen that 
$\{0_a,0_{\sigma (a)}\}$, $\sigma (a)\neq a$ is part of the root configuration consistent with the Bethe equations. These accidental zeros are not fixed points of parity and do not require special treatment.
When talking about zero roots in what follows we specifically mean unpaired roots on the
nodes invariant under parity transformation. 
A parity-invariant state can thus be  characterized by half of the paired roots $u_{aj}$ and the nodes $a_\alpha $ hosting solitary zero roots\footnote{In the $SU(4)$ case solitary zero roots can only sit on the middle node. We prefer to keep the discussion general  in view of  eventual generalization to the full superconformal group.}. 

Overlaps of  integrable boundary states with Bethe eigenstates are described by a remarkably compact formula, which is universal to a large degree albeit its detailed structure is model-specific and is actually known on a case-by-case basis. A systematic derivation of the overlap formulas goes back to the work of Tsuchiya \cite{Tsuchiya:qf} and has so far been restricted to $\mathfrak{su}(2)$ \cite{Brockmann:2014a,Brockmann:2014b} (see also~\cite{Buhl-Mortensen:2015gfd,Foda:2015nfk}). 
Known overlaps for higher-rank groups are based on educated guesswork aided by symmetry consideration \cite{deLeeuw:2016umh,DeLeeuw:2018cal,deLeeuw:2019ebw}, while a systematic derivation has only appeared recently for a class of $SU(N)$ spin chains \cite{Gombor:2021hmj}. Fortunately, the $SU(4)$ spin chain of ABJM is precisely of the type considered in  \cite{Gombor:2021hmj}. In fact, the overlap for the simplest case of VBS  (\ref{MPS->VBS}), (\ref{2-block}) is  
explicitly given in \cite{Gombor:2021hmj}\footnote{It was conjectured in  \cite{Gombor:2021hmj} that the overlap formula derived there should describe domain walls in  ABJM theory. This conjecture turns out to be correct. In fact,
the overlap formula in \cite{Gombor:2021hmj} applies to a larger class of valence-bond states that form a one-parameter family with the two-site matrix $\mathcal{K}(a)=1+a\sigma _1\otimes \sigma _1$ which for $a=1$ is related to (\ref{2-block}) by
an $SU(4)$ transformation: $K=S^\dagger U^\dagger \mathcal{K}(1)US$ with $U=(1-i\sigma _2)\otimes(1-i\sigma _2)/2$ and $S$ representing the $(24)$ permutation. It would be interesting to understand if the domain-wall solution admits a one-parameter deformation that corresponds to the boundary states with $a\neq 1$. The two-site matrix $\mathcal{K}(a)$ is non-degenerate for $a\neq \pm 1$ and the putative domain wall solution should involve all four scalars. The string dual is then expected to be a D8 brane wrapping the whole $\mathbb{C}P^3$. Candidate brane configurations have been considered in \cite{Fujita:2009kw}, and it would be interesting to find their field-theory counterparts.}:
\begin{equation}\label{overlap}
 \frac{\left\langle {\rm VBS}\right.\!\left|\mathbf{u} \right\rangle}{\left\langle \mathbf{u}\right.\!\left|\mathbf{u} \right\rangle^{\frac{1}{2}}}
 =2^{-L}\,{Q_2(i)}\sqrt{\frac{\mathop{\mathrm{Sdet}}G}{Q_2(0)Q_2\left(\frac{i}{2}\right)}}
 \,.
\end{equation}

The key ingredient of the overlap formula is the superdeterminant of the Gaudin matrix
\begin{equation}
 G_{aj,bk}=\frac{\partial \chi _{aj}}{\partial u_{bk}}\,,
\end{equation}
where the $\mathbbm{Z}_2$ grading of the superdeterminant is defined  by the parity transformation (\ref{parity}) acting on rows and columns of the Gaudin matrix. In the basis where parity acts diagonally the superdeterminant becomes a ratio of ordinary determinants:
\begin{equation}
 \mathop{\mathrm{Sdet}}G=\frac{\det G^+}{\det G^-}\,.
\end{equation}

The rows and columns of $G^-$ are labeled by positive Bethe roots only, while $G^+$  includes in addition the zero roots:
\begin{eqnarray}\label{G+-}
 G^\pm_{aj,bk}&=&\left(\frac{Lq_a}{u_{aj}^2+\frac{q_a^2}{4}}
 -\sum_{cl}^{}K^+_{aj,cl}-\frac{1}{2}\sum_{\alpha }^{}K^+_{aj,a_\alpha 0}
 \right)\delta _{ab}\delta _{jk}+K^\pm_{aj,bk},
\nonumber \\
G^+_{aj,\alpha }&=&\frac{1}{\sqrt{2}}\,K^+_{aj,a_\alpha 0},
\nonumber \\
G^+_{\alpha \beta }&=&
\left(\frac{4L}{q_{q_\alpha }}-\sum_{cl}^{}K^+_{a_\alpha 0,cl}
-\sum_{\gamma }^{}\frac{4}{M_{a_\alpha a_\gamma }}\right)
\delta _{\alpha \beta }
+\frac{4}{M_{a_\alpha a_\beta }}\,,
\end{eqnarray}
where
\begin{equation}
 K^\pm_{aj,bk}=\frac{M_{ab}}{\left(u_{aj}-u_{bk}\right)^2+\frac{M_{ab}^2}{4}}
 \pm
 \frac{M_{a\sigma (b)}}{\left(u_{aj}+u_{bk}\right)^2+\frac{M_{a\sigma (b)}^2}{4}}\,.
\end{equation}
Potentially ill-defined quantities, like $1/M_{ab}$ for $M_{ab}=0$, are declared to be zero.

The Gaudin factor in the overlap formula is decorated by the Baxter polynomials:
\begin{equation}
 \Q_a(u)=\prod_{{\rm all~}a{\rm~roots}}^{}\left(u-u_{aj}\right).
\end{equation}
To be more precise, the Q-functions in the overlap formula are regularized by omitting eventual zero roots. Hopefully the use of the same notations for the full and regularized Baxter polynomials will not cause any confusion.

Together with (\ref{O(x)}), the overlap formula (\ref{overlap}) describes the tree-level expectation values of all scalar operators\footnote{As it stands, the formula applies to conformal primaries. Descendants acquire an extra group-theoretic factor \cite{deLeeuw:2017dkd,Gombor:2021uxz},  which we do not display here.}. Before extending the overlap formula to the full set of operators, we consider a few representative examples.

As discussed in sec.~\ref{selection},
primaries with non-zero overlaps must be $SU(4)$ singlets. There are 
two singlets at length two, the simplest possible case. The Bethe-ansatz description of one of them is singular, involving roots at $\pm i/2$ \cite{Minahan:2008hf}. The Gaudin matrix then has to be regularized and redefined. The correct prescription of absorbing infinities is known  \cite{Kristjansen:2021xno}, but its detailed outline would create unnecessary complications and since our goals are mostly illustrative, we concentrate on the other operator
\begin{equation}\label{O2}
 \mathcal{O}_2=\mathop{\mathrm{tr}}\left(
  Y^AY^\dagger _AY^BY^\dagger _B
 +Y^AY^\dagger _BY^BY^\dagger _A
 \right),
\end{equation}
having energy $10\lambda ^2$ and being described by regular Bethe roots \cite{Minahan:2008hf}:
\begin{equation}
 u_1=\left\{\sqrt{\frac{3}{20}}\,,-\sqrt{\frac{3}{20}}\right\}=u_3,
 \qquad 
 u_2=\left\{\frac{1}{\sqrt{5}}\,,-\frac{1}{\sqrt{5}}\right\}.
\end{equation}

Taking the positive roots, one per node, to represent the state\footnote{We can equally well represent pairs with $\{\{u,-u\},\{v\},\{\}\}$. This does not change the result.}:
 $$u_{aj}=\{\{u\},\{v\},\{u\}\},$$ 
 we get for the Gaudin factors defined in (\ref{G+-}):
\begin{equation}
 G^\pm=\begin{bmatrix}
  \frac{3}{2}\,\frac{1}{\frac{1}{4}+u^2} +\kappa^+  & -\kappa ^\pm  &  \pm \frac{1}{2}\,\frac{1}{\frac{1}{4}+u^2}  \\ 
 -\kappa ^\pm  & 2\kappa ^+-\frac{1}{2}\,\frac{1\mp 1}{\frac{1}{4}+v^2}  &  -\kappa ^\pm \\ 
  \pm \frac{1}{2}\,\frac{1}{\frac{1}{4}+u^2}  & -\kappa^\pm  & \frac{3}{2}\,\frac{1}{\frac{1}{4}+u^2}+\kappa ^+  \\ 
 \end{bmatrix},
\end{equation}
where
\begin{equation}
 \kappa ^\pm=\frac{1}{\frac{1}{4}+(u-v)^2}\pm \frac{1}{\frac{1}{4}+(u+v)^2}.
\end{equation}
Setting $u=\sqrt{3/20}$, $v=1/\sqrt{5}$, and calculating the determinants we find:
\begin{equation}
 \mathop{\mathrm{Sdet}}G\equiv \frac{\det G^+}{\det G^-}=\frac{9}{10}\,.
\end{equation}
The one-point function follows from (\ref{overlap}), (\ref{O(x)}):
\begin{equation}
 \left\langle \mathcal{O}_2(x)\right\rangle=\frac{1}{x^2}\,
 \frac{3}{2\sqrt{5}\,\lambda ^2}\,.
\end{equation}
Properly normalizing the operator and replacing the fields in (\ref{O2}) by the classical solution gives, of course, the same result.

Another example is a length-three operator
\begin{eqnarray}
 \mathcal{O}_3&=&\left(\sqrt{10}-1\right)\mathop{\mathrm{tr}}\left(
 Y^AY^\dagger _AY^BY^\dagger _BY^CY^\dagger _C
 +Y^AY^\dagger _CY^BY^\dagger _AY^CY^\dagger _B
 \right.
\nonumber \\
&&\left.
 -Y^AY^\dagger _BY^BY^\dagger _CY^CY^\dagger _A
 \right)
 +9\mathop{\mathrm{tr}}Y^AY^\dagger _AY^BY^\dagger _CY^CY^\dagger _B,
\end{eqnarray}
with the anomalous dimension $(8+2\sqrt{10})\lambda ^2$ and root content \begin{equation}
 u_1=\left\{\alpha ,-\alpha ,0\right\}=u_3,\qquad 
 u_2=\left\{\frac{2\alpha }{\sqrt{3}}\,,-\frac{2\alpha }{\sqrt{3}}\,,0\right\},\qquad 
 \alpha^2 =\sqrt{\frac{2}{5}}-\frac{1}{4}\,.
\end{equation}
It is easy to check that the operator is an eigenstate of the Hamiltonian (\ref{1-loop-mixing}) and that the root configuration solves the Bethe equations.

The zero roots on nodes $1$ and $3$ are accidental, they are paired with one another. 
On the contrary, the zero on node $2$ is solitary and does not have a pair.  
Because of that the Gaudin factors have different sizes, $5\times 5$ for $G^+$ and $4\times 4$ for $G^-$, see (\ref{G+-}). 
We can represent pairs by $\{\{\alpha ,-\alpha ,0\},\{2\alpha /\sqrt{3}\},\{\}\}$,
for example.
Evaluating the determinants we find:
\begin{equation}
 \left\langle \mathcal{O}_3(x)\right\rangle=\frac{1}{x^3}\,\,\frac{1}{\lambda ^3}
 \sqrt{\frac{54+17\sqrt{10}}{390}}\,.
\end{equation}
We checked that replacement of scalar fields in the operator by the domain wall solution gives the same result.

\subsection{Full spectrum \label{Full}}

\begin{figure}[t]
\begin{center} 
 \subfigure[]{
   \includegraphics[height=3.6cm] {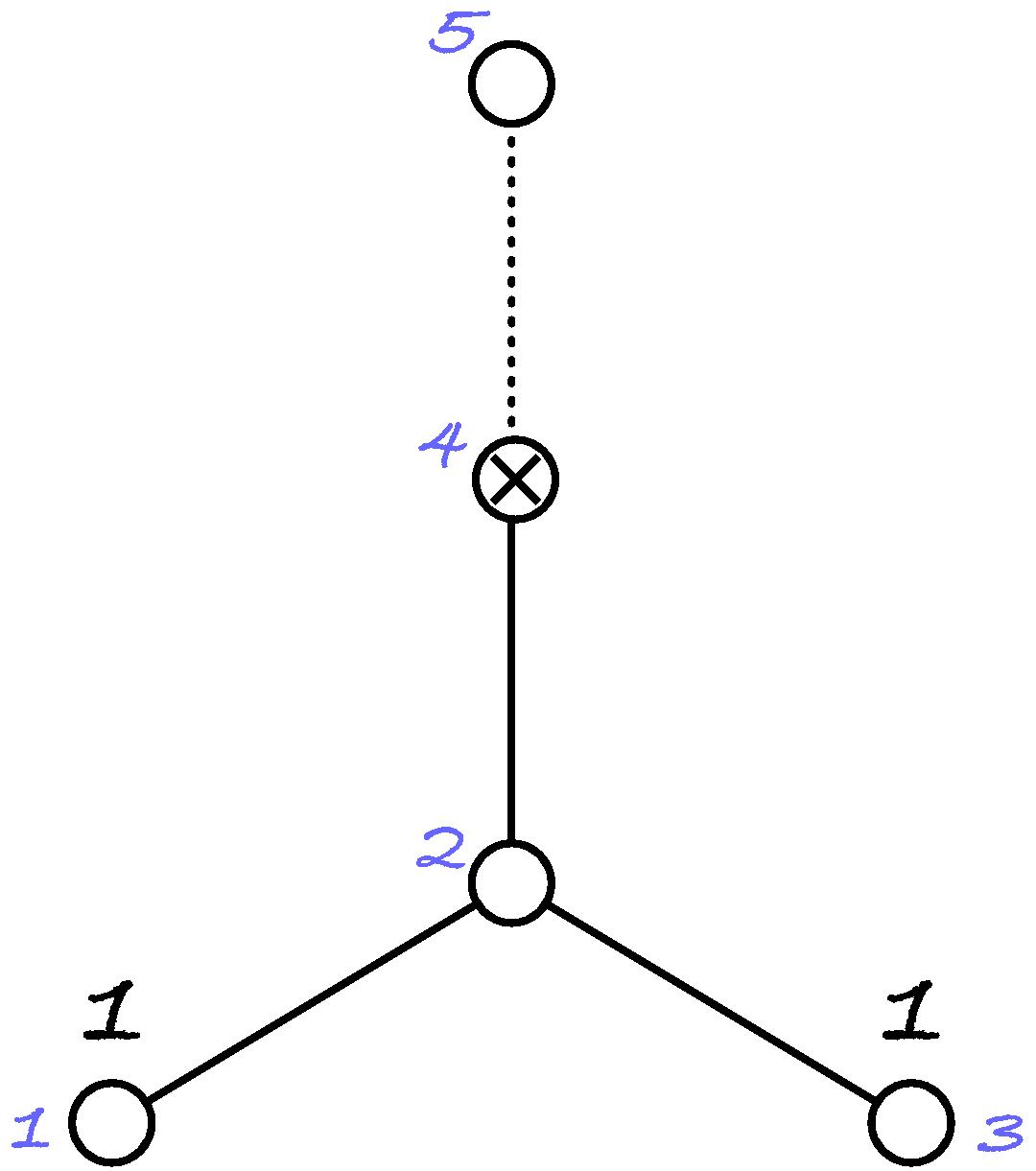}
   \label{Dynkin-SU4}
 }
 \subfigure[]{
   \includegraphics[height=3.6cm] {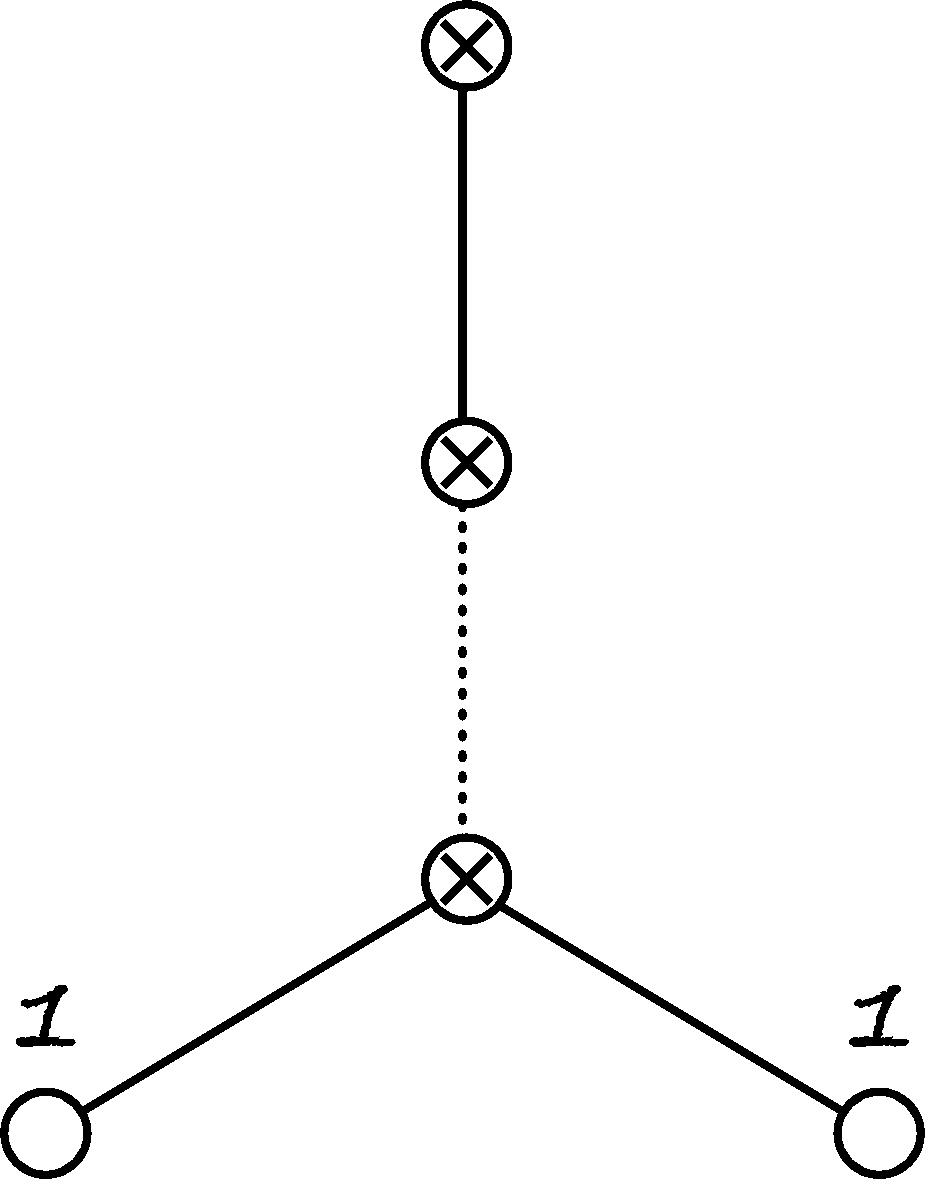}
   \label{Dynkin-F}
 }
 \subfigure[]{
   \includegraphics[height=3.6cm] {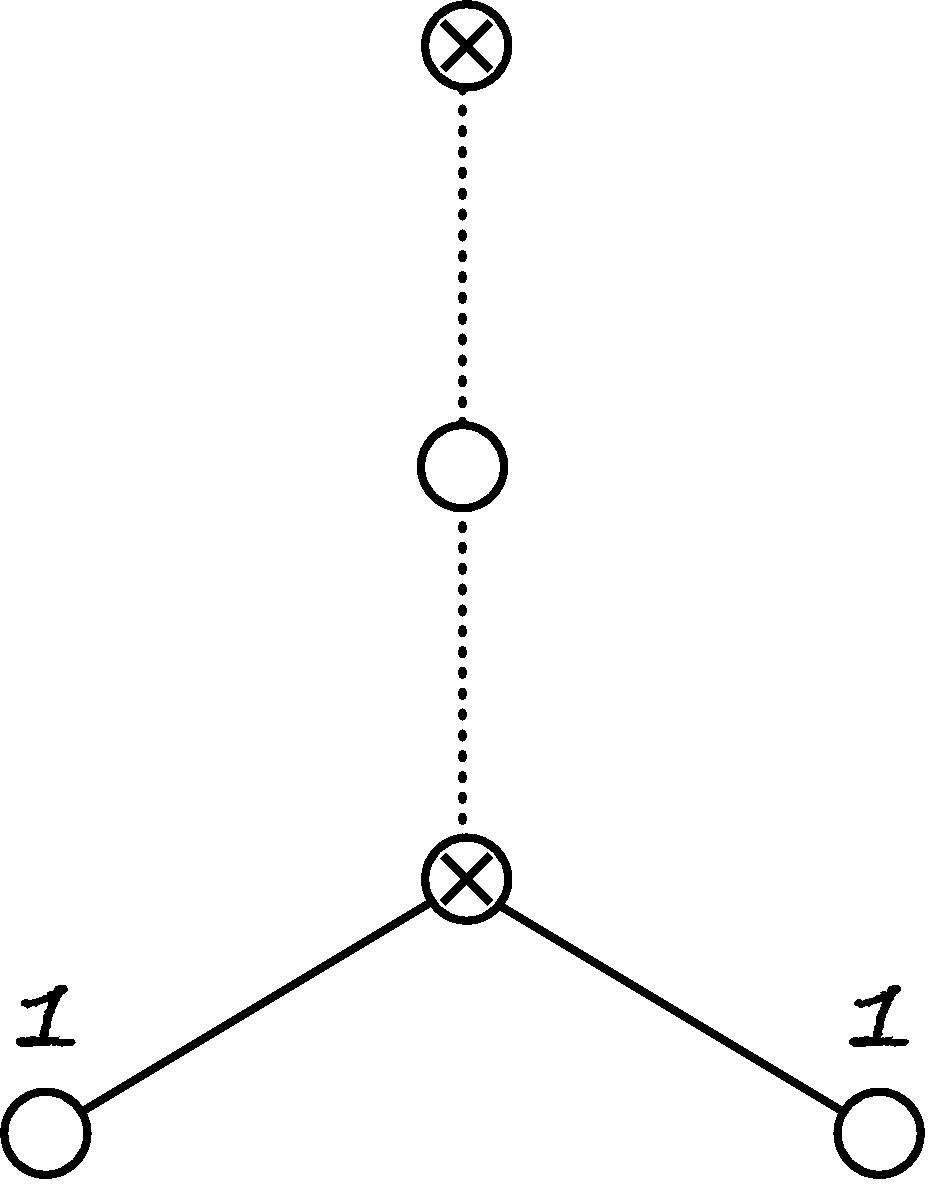}
   \label{Dynkin-D}
 }
 \subfigure[]{
   \includegraphics[height=3.6cm] {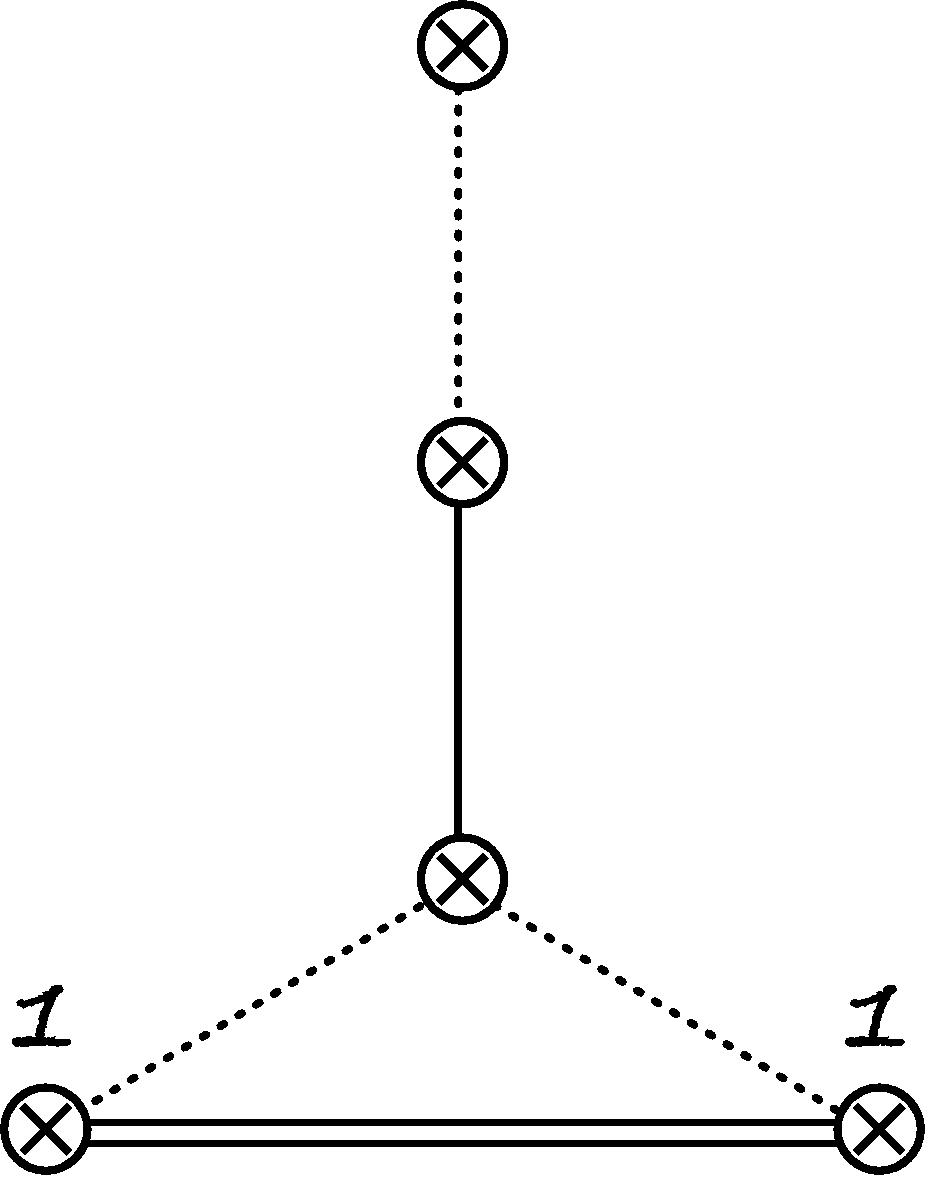}
   \label{Dynkin-ML}
 }
\caption{\label{Dynkin-diagrams}\small 
Dynkin diagram of $\mathfrak{osp}(6|4)$ in different gradings.
}
\end{center}
\end{figure}

All possible operators in the ABJM theory are described by the $OSp(6|4)$ Bethe equations. The Dynkin diagram is shown in  fig.~\ref{Dynkin-SU4}, where as usual the matrix element for a solid link is $-1$, $+1$ for a broken link, $0$ for a fermionic (crossed) node, $+2$ for a bosonic node between solid links and $-2$ between broken links. The spin representation has non-zero Dynkin indices on the nodes 1 and 3 as shown in the figure. Nodes 1, 2 and 3 correspond to the scalar $SU(4)$ subsector discussed so far. An excitation on node 4 transforms a boson into a fermion and that on node 5 executes a spin flip.

\subsubsection{Fermionic dualities}

Since a Dynkin diagram for a superalgebra is not unique \cite{Frappat:1996pb}, the Bethe equations for a super-spin chain can be written in several different forms related by fermionic dualities \cite{Tsuboi:1998ne}. For an example, consider the fermionic node in the original $SU(4)$-friendly diagram~\ref{Dynkin-SU4}. The Bethe equations for the fermionic roots are
\begin{equation}\label{fBethe}
 1=\left.\frac{\Q^+_5\Q^-_2}{\Q^-_5\Q^+_2}\right|_{u_{4j}}.
\end{equation}
They can be equivalently represented as a QQ-relation between the Baxter polynomials:
\begin{equation}\label{QQ}
 \Q^-_5\Q^+_2-\Q^+_5\Q^-_2=i(K_2-K_5)\Q_4\widetilde{\Q}_4,
\end{equation}
where  $\Q_4$ is the original Q-function and  $\widetilde{\Q}_4$ is its complement that absorbes the remaining $K_2+K_5-K_4-1$ roots of the polynomial of the left-hand side. Here we use the standard notations for the argument shifts
\begin{equation}
 f^{[q]}(u)=f\left(u+\frac{iq}{2}\right),\qquad f^\pm=f^{[\pm 1]},\qquad f^{\pm\pm}=f^{[\pm 2]}.
\end{equation}

While the Bethe equation (\ref{fBethe}) holds on the fermionic roots, the QQ-relation (\ref{QQ}) is a functional equation valid for arbitrary value of the argument in $Q_a(u)$. It is easy to see that the roots of $\widetilde{\Q}_4$, which we denote by $\widetilde{u}_{4j}$, satisfy the same Bethe equations as $u_{4j}$. Moreover, the scattering phases on the adjacent nodes can be expressed through $\widetilde{u}_{4j}$ with the help of (\ref{QQ}):
\begin{equation}
 \left.\frac{Q^+_4}{Q^-_4}=-\frac{Q^{++}_{2,5}\widetilde{Q}^-_4}{Q^{--}_{2,5}\widetilde{Q}^+_4}\right|_{u_{2,5j}}.
 \end{equation}
The $Q_{2,5}^{\pm\pm}$ factors neatly cancel in the Bethe equations and we arrive at a new system associated with the Dynkin diagram in fig.~\ref{Dynkin-F}.

Dualizing further on node $5$ we get the diagram in fig.~\ref{Dynkin-D}, in many respects  distinguished. It is this diagram that allows for an extension to all loop orders  \cite{Gromov:2008qe}. 

The next step gives rise to a new feature. The QQ-relation on the  central node contains a product of three Q-functions:
\begin{equation}\label{QQQ}
 Q^-_4Q^+_1Q^+_3-Q^+_4Q^-_1Q^-_3=i(K_1+K_3-K_4)Q_2\widetilde{Q}_2.
\end{equation}
As a consequence, the scattering phase in the Bethe equations for roots 1 and 3 contains an extra piece:
\begin{equation}
 \left.\frac{Q^+_2}{Q^-_2}=-\frac{Q^{++}_1Q^{++}_3\widetilde{Q}^-_2}{Q^{--}_1Q^{--}_3\widetilde{Q}^+_2}\right|_{u_{1,3j}}\,.
\end{equation}
Substituting this into the Bethe equations for roots 1 and 3 not only cancels their self-scattering, but induces a mutual interaction depicted as a double line in fig.~\ref{Dynkin-ML}. The Dynkin diagram ceases to be simply-laced.

\subsubsection{$OSp(6|4)$ overlaps}

The Gaudin superdeterminant transforms under fermionic dualities in a regular way, such that determinants constructed on the original and dual roots are related by a simple Jacobian. Namely, for the duality transformation on the $a$-th node, the following relation holds\footnote{The formula seems to apply to simply-laced Dynkin diagrams, and may need to be modified in the non-simply-laced case, for example when dualizing on one of the nodes in fig.~\ref{Dynkin-ML} connected by the double line.}:
\begin{equation}\label{duality-overlap}
 \prod_{b:~M_{ab}\neq 0}^{}Q_b(i/2)\,\mathop{\mathrm{Sdet}}\widetilde{G}=Q_a(0)\widetilde{Q}_a(0)\,\mathop{\mathrm{Sdet}}G,
\end{equation}
where the product on the left-hand side is over all nodes adjacent to $a$. This formula was conjectured for $A_n$-type Dynkin diagrams and checked for a number of chiral boundary states \cite{Kristjansen:2020vbe,Kristjansen:2021xno}. The boundary state at hand is achiral and the Dynkin diagram is of the $D$-type, but we found that the formula continues to hold without change. Although we have no mathematical proof, we performed thorough numerical checks for all duality transformations considered in the previous subsection. 

The transformation law of the Gaudin superdeterminant can be used to transform 
the overlap formula from one duality frame to another.
The  overlap of an integrable boundary state is expected to have a universal form
\begin{equation}\label{basic-block}
 \sqrt{\prod_{a}^{}\frac{\prod_{j}^{}Q_a(i\alpha _{aj}/2)}{\prod_{k}^{}Q_a(i\beta _{ak}/2)}\,\mathop{\mathrm{Sdet}G}}\,,
\end{equation}
or to be a linear combination of such terms. Ultimately, this structure is dictated by the underlying Algebraic Bethe Ansatz of rational type \cite{Gombor:2021hmj,Buhl-Mortensen:2015gfd}. But we simply accept this formula as an  assumption. By itself this is not very constraining, but in conjunction with fermionic duality this structural assumption places powerful constraints on the particular values of the Q-functions that can appear.

\begin{figure}[t]
 \centerline{\includegraphics[width=4cm]{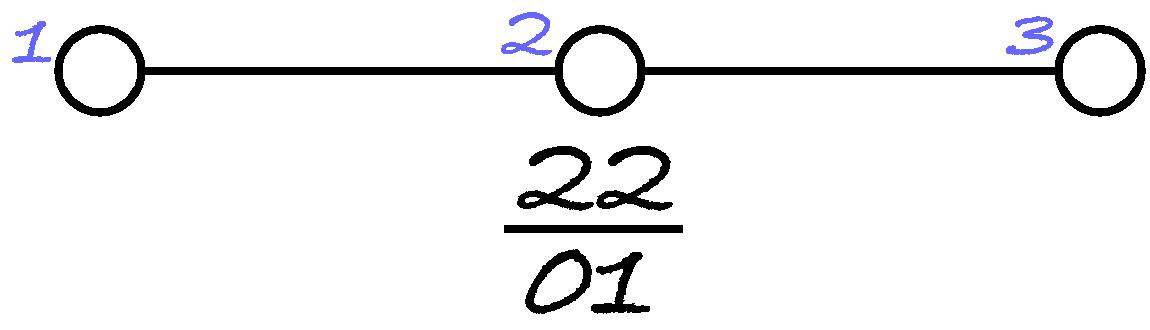}}
\caption{\label{su4}\small The $SU(4)$ overlap formula in the graphic notations.}
\end{figure}

We use a graphic notation introduced in \cite{Kristjansen:2020vbe} for the basic building block (\ref{basic-block}) by placing numbers $\alpha _{aj}$, $\beta _{ak}$ directly on the Dynkin diagram:
\begin{equation}
 \includegraphics[height=1.5cm] {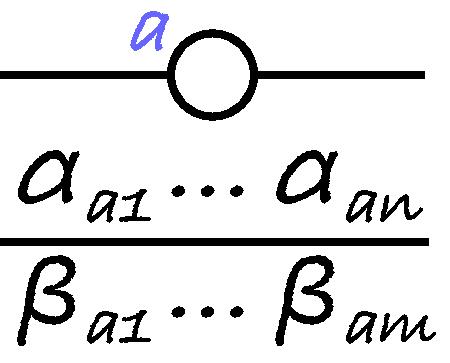}
\end{equation}
For example, the overlap formula (\ref{overlap}) for the $SU(4)$ sectors is represented by fig.~\ref{su4}.

It is not hard to see that an extension to the full $OSp(6|4)$ is actually unique as long as it respects fermionic duality. The transformation law (\ref{duality-overlap}) leaves two possibilities for a fermionic node, ${0}/{}$ or ${}/{0}$, if the overlap is to retain its structure in the new frame. Otherwise the dependence on the original fermionic roots, that are traded for their duals, would not completely cancel. For the particular case at hand $0/$  is preferable, the Jacobian then cancels $/1$ on the 2nd node, and that is going to help in the subsequent steps. Since the 5th node becomes fermionic as well, the only choice there is $0/1$. The $/1$ is neatly cancelled by the Jacobian leaving $0/$ which can again be dualized.

\begin{figure}[t]
 \centerline{\includegraphics[width=11cm]{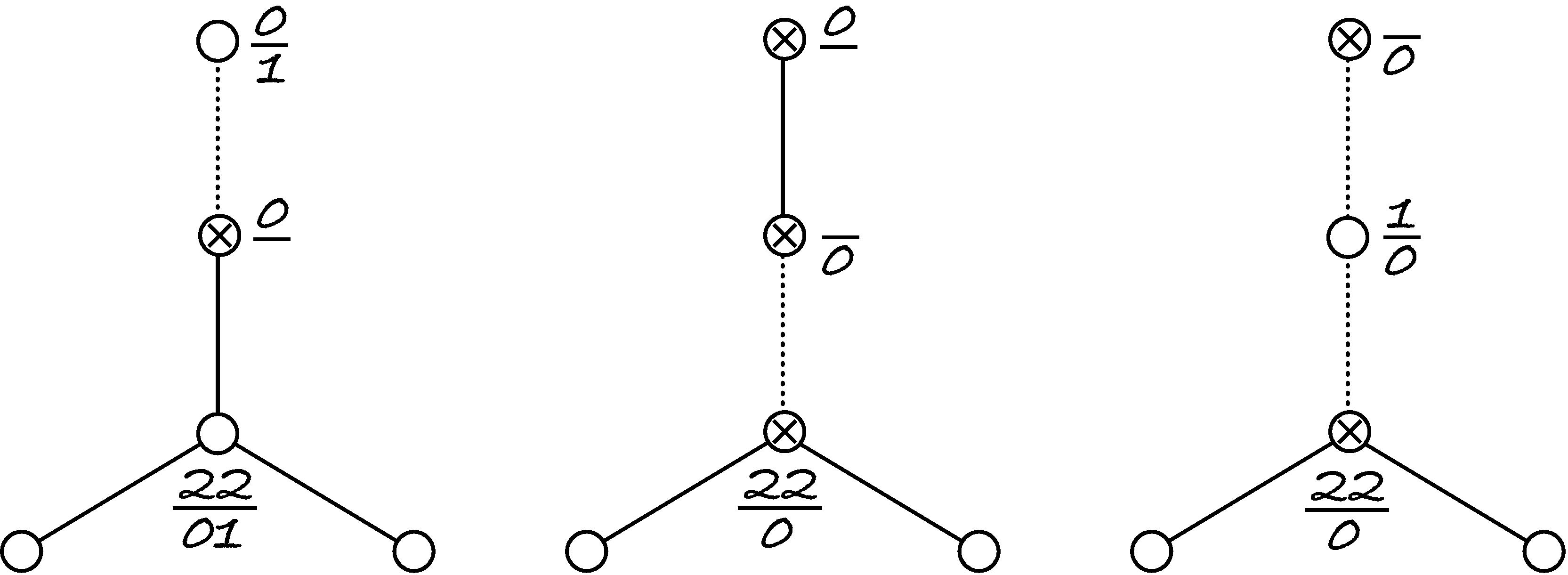}}
\caption{\label{allos}\small Overlap formulas in different grading.}
\end{figure}

Arguments of this kind reconstruct overlaps for 
the first three diagrams in fig.~\ref{Dynkin-diagrams} with the result shown in fig.~\ref{allos}. Duality on the 4th node of the first diagram replaces $Q_4(0)\mathop{\mathrm{Sdet}}G$ by $Q_2(i/2)Q_5(i/2)\mathop{\mathrm{Sdet}}\widetilde{G}/\widetilde{Q}_4(0)$ which gives the second diagram. Likewise,  dualizing the 5th node of the second diagram replaces $Q_5(0)\mathop{\mathrm{Sdet}}G$ by $Q_4(i/2)\mathop{\mathrm{Sdet}}\widetilde{G}/\widetilde{Q}_5(0)$ yielding the third diagram.

This diagram is of particular interest,  because its Bethe equations can be extended to all orders in the coupling. The explicit form of the overlap for this distinguished grading is
\begin{equation}
 \frac{\left\langle {\rm MPS}_2\right.\!\left| \mathbf{u}\right\rangle}{\left\langle \mathbf{u}\right.\!\left| \mathbf{u}\right\rangle^{\frac{1}{2}}}
 =Q_2(i)\sqrt{\frac{Q_4(i/2)}{Q_2(0)Q_4(0)Q_5(0)}\,\mathop{\mathrm{Sdet}}G}\,.
\end{equation}
We expect this formula to have an all-loop generalization, but we will not attempt to derive or guess it here.

\begin{figure}[t]
 \centerline{\includegraphics[width=9cm]{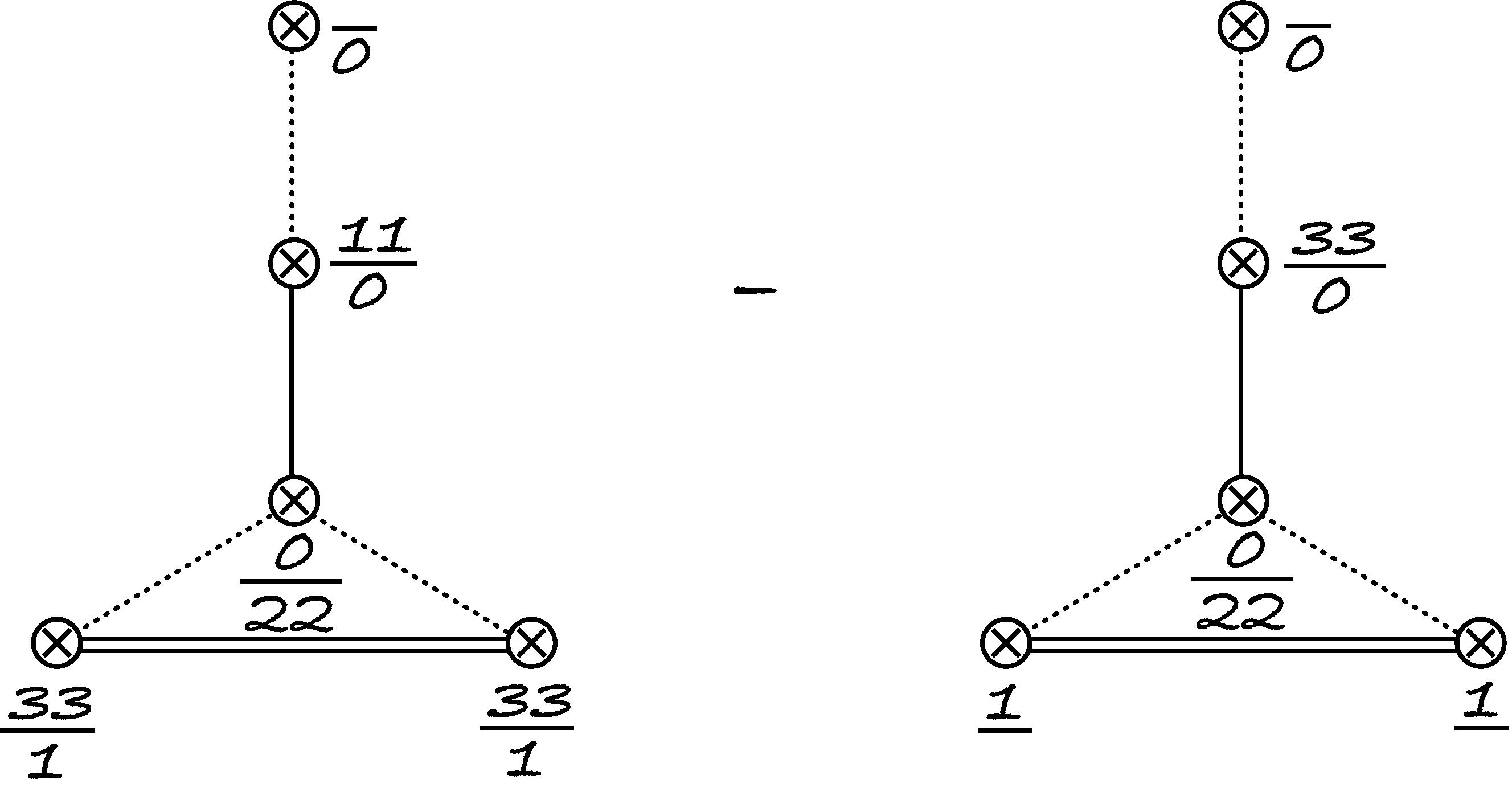}}
\caption{\label{MLo}\small Overlap formula for a non-simply-laced form of the Dynkin diagram.}
\end{figure}

It is also possible to dualize on the central node, albeit the outcome is somewhat different compared to the simply-laced case. The factor of $Q_2(i)$ is not accounted for by the Jacobian and has to be expressed through the other Q-functions\footnote{The duality transformation must act from the diagram in~\ref{Dynkin-ML} to \ref{Dynkin-D}, so as to replace $Q_2(0)$ by $1/\widetilde{Q}_2(0)$, and we need to express the dual Q-function through the original ones.}, which is possible in virtue of the QQ-relation (\ref{QQQ}):
\begin{equation}
 \widetilde{Q}_2(i)\propto\frac{Q_4(i/2)Q_1(3i/2)Q_3(3i/2)}{{Q}_2(i)}-
 \frac{Q_4(3i/2)Q_1(i/2)Q_3(i/2)}{{Q}_2(i)}\,.
\end{equation}
Standing out of the square root this factor does  not compromise the general structure.
The dualized overlap becomes a difference of two terms each of the canonical form, as illustrated in fig.~\ref{MLo}.

\section{One-point functions at strong coupling \label{fluctuations}}

One-point functions at strong coupling can be calculated by a variant of the GKPW prescription~\cite{Gubser:1998bc,Witten:1998qj} which entails a fluctuation
analysis of the  supergravity background. The foundation for the fluctuation analysis for type IIA supergravity on $AdS_4 \times S^7$ was laid in~\cite{Biran:1983iy,Castellani:1984vv,Bastianelli:1999bm} and employed in the study
 the study three-point functions involving two giant gravitons and one tiny graviton in ABJM theory 
 in~\cite{Hirano:2012vz,Yang:2021kot} as well as in the computation 
of correlation functions of vortex loop operators with local operators in~\cite{Drukker:2008jm}. 

Hence, let us  consider the variation of the 
Euclidean DBI and WZ action.
For the latter we have
\begin{equation}
\delta{I_{WZ}}= T_4 \int d^5 \sigma\,  {\cal F} \wedge {\cal P}[\delta C_3],
\end{equation}
where
\begin{equation}
(\delta C_3)_{\mu_1 \mu_2 \mu_3}= \frac{R^3}{8} \, 2 \epsilon_{\mu_1 \mu_2 \mu_3 \mu_4} \,\nabla^{\mu_4} s^{\Delta I}(X)
Y_{\Delta I}(\Omega),
\end{equation}
with the $\mu$'s referring to coordinates on $AdS_4$ and $\epsilon_{1 2 3 4}=\sqrt{g_{AdS_4}}=z^2$.
  The $X$ likewise refers to $AdS_4$ coordinates while $\Omega$ 
refers to coordinates on $\mathbb{C}P^3$. For the relevant pull-back to the brane world-volume we get
\begin{eqnarray}
{\cal P}[\delta C_3]_{0 1 z}&=& \frac{R^3}{8} 2 \left( x_2'(z)  \epsilon_{012z} \, z^2 \partial_z
+    \epsilon_{01z2}\frac{1}{z^2} \partial_{x_2} 
\right) s^{\Delta I}(X) Y_{\Delta I }(\Omega)\\
&=& \frac{R^3}{8} 2 z^2 \left( -Q \, \partial_z
-\frac{1}{z^2} \partial_{x_2} 
\right) s^{\Delta I }(X) Y_{\Delta I}(\Omega).
\end{eqnarray}
We can now combine this with ${\cal F}$ and get
\begin{eqnarray}
\delta I_{WZ}&=&T_4 \widetilde{R}^4  \sqrt{\alpha'}\,k\, Q\int_0^{\infty} \, dz\, z^2\, \int_{-\infty}^{\infty} dx_0 \, dx_1\, 
\int_0^\pi d\theta_1  \sin\theta_1\int_0^\pi d\phi_1 \\
&& \times\left( -Q \, \partial_z
-\frac{1}{z^2} \partial_{x_2} 
\right) s^{\Delta I}(X) Y_{\Delta I}(\Omega). \label{deltaWZ} \nonumber
\end{eqnarray}
Moving on to the DBI action we find for the variation
\begin{equation}
\delta I_{DBI}=-T_4 e^{-\Phi} \frac{1}{2} \int d^5 \sigma \sqrt{\det (G+{\cal F})} \,(G+{\cal F})^{ab} \delta G_{ab},
\end{equation}
where for the $AdS$ part of the metric we have
\begin{equation}
\delta g_{\mu \nu}=\frac{4}{\Delta+2} \left[ \nabla_{\mu} \nabla_{\nu}-\frac{1}{6}\Delta (\Delta-1)\,g_{\mu \nu}\right]
s^{\Delta I}(X) Y_{\Delta I}(\Omega),
\end{equation}
whereas on $\mathbb{C}P^3$
\begin{equation}
\delta g_{\alpha \beta}=\frac{\Delta}{3} \,g_{\alpha \beta} \, s^{\Delta I}(X) Y_{\Delta I}(\Omega).
\end{equation} 
To compute the double covariant derivative we will need some Christoffel symbols for $AdS_4$. The relevant ones
are
\begin{equation}
 \Gamma_{00}^z=\Gamma_{11}^z=\Gamma_{22}^z= -z^3, \hspace{0.5cm}
 \Gamma_{zz}^z=-\frac{1}{z}, \hspace{0.5cm}
\Gamma_{z2}^z=\Gamma_{2z}^z=\frac{1}{z}, 
\end{equation}
For the calculation of the variation we also need the inverse of the induced metric. Its non-zero relevant elements are
\begin{eqnarray}
(G+{\cal F})^{00}&=&(G+{\cal F})^{11}=\frac{1}{z^2}, \hspace{0.5cm} (G+{\cal F})^{zz}= 
\left(\frac{1}{z^2}+z^2(x_2'(z))^2\right)^{-1}= \frac{z^2}{1+Q^2},\nonumber \\
(G+{\cal F})^{\theta_1 \theta_1}& =&\frac{1}{1+Q^2}, \hspace{0.5cm} (G+{\cal F})^{\phi_1 \phi_1} =
\frac{1}{\sin^2\theta_1(1+Q^2)},
\end{eqnarray}
where we have set $\cos^2 \xi=1$.
Of the fluctuations, only $\delta G_{zz}$ is a bit complicated involving a non-trivial pull back. 
 From eqn.~(\ref{inducedmetric}) we get
\begin{equation}
\sqrt{\det (G+{\cal F})}=\widetilde{R}^{5}\,z   \sin \theta_1 \,(1+Q^2),
\end{equation}
Putting everything together, we find
\begin{eqnarray} \label{DeltaDBI}
\delta I_{DBI} &=&-T_4 \widetilde{R}^4 k \,\sqrt{\alpha'} (1+Q^2) \int_0^{\infty}dz\, z\int_{-\infty}^{\infty} dx_0 \, dx_1 \
\int_0^{2\pi} d\phi_1 \int_0^\pi \ d\theta_1 \sin \theta_1\nonumber\\
&&\times \left\{
\frac{1}{\Delta+2} \left[ \nonumber
\frac{1}{z^2} \left(\partial_0^2+\partial_1^2 \right) + 3 z \partial_z \right. \right.\\
&&+\left.\frac{z^2}{1+Q^2}\left( \partial_z^2-2\frac{Q}{z^2} \left (\partial_z\partial_2 -\frac{1}{z}\partial_2\right)+\frac{Q^2}{z^4} 
\partial_2^2
\right) -\frac{1}{2}\Delta (\Delta-1)\nonumber
\right] \\ 
&& \left.+\frac{\Delta}{6}\frac{1}{1+Q^2} \right\} 
s^{\Delta I}(X)
Y_{\Delta I}(\Omega).
\end{eqnarray}
 We note that
 the term in the last line originates from the spherical part of the geometry and the remaining terms come from
the $AdS$ part.

Implementing the GKPW prescription, i.e.\ differentiating after a delta-function source on the boundary coupling to a specific chiral primary CFT operator  and setting the source to zero corresponds to picking out the term involving the corresponding spherical harmonic and replacing the mode $s_{\Delta I }(X)$ by its bulk to boundary propagator, i.e.
\begin{equation}\label{replacement}
\left. \langle  {\cal O}({x})_{\Delta}\rangle = -\delta I_{DBI+WZ} \right| s \to  \frac{C_{\Delta}}{z^\Delta\left(\rho^2 + (x_2-{x})^2 +1/z^2\right)^\Delta},
\hspace{0.7cm} \rho^2=x_0^2+x_1^2, \nonumber
\end{equation}
where (see f.inst.~\cite{Drukker:2008jm})
\begin{equation} \label{CDelta}
C_{\Delta}=
 \sqrt{k}\, \left(\frac{l_p}{R}\right)^{9/2}{\cal C}_{\Delta},
\end{equation}
with
\begin{equation}
{\cal C}_{\Delta}=2^{\Delta/2-1}\,  \pi  \,\,\frac{\Delta+2}{\Delta} \sqrt{\Delta+1}.
\end{equation}
Next step is to perform all the integrations and differentiations implied by the relations~(\ref{deltaWZ}) and~(\ref{DeltaDBI}).
Here one can check that the term with $\partial_0^2+\partial_1^2$ does not give any contribution and we can rewrite
$\int_{-\infty}^{\infty} dx_0 dx_1=2\pi \int_0^\infty \rho\, d\rho$.

Furthermore,
we notice that the common pre-factor of the two integrals can be rewritten as
\begin{equation}\label{prefactor}
T_4\widetilde{R}^4 k  \sqrt{\alpha'}= \frac{N}{8\pi^2}.
\end{equation}

From~(\ref{deltaWZ}) and~(\ref{DeltaDBI}) we immediately see that the standard chiral primary $(Y_1 {Y}^{*}_2)^L$
has a vanishing one-point function as its evaluation involves the integral $\int_0^{2\pi} d\phi_1\exp(i\phi_1)=0$. On the other
hand the chiral primaries defined by the relation~(\ref{Laplace}) have non-vanishing one-point functions.

Combining the factors~(\ref{CDelta}) and~(\ref{prefactor}) we see that in the strong coupling limit all one-point functions will carry a pre-factor  
\begin{equation}
\sim \left(\frac{l_p}{R}\right)^{9/2} \sqrt{k} N \sim k^{-3/2} \,\lambda^{-3/4} \sqrt{k} N~\sim \lambda^{1/4},
\end{equation}
and since the integrals when expanded for large $Q$ involve only integer powers of $Q$
we will not immediately be able to compare to a perturbative gauge theory
calculation even in the double scaling limit~(\ref{dsl}).  This is in contrast to the $AdS_5\times S^5$ case where the strong 
coupling expansion in combination with a double scaling limit gave rise to one point functions 
expandable in integer powers of $\lambda$~\cite{Nagasaki:2012re,Kristjansen:2012tn}.

Let us now calculate the one-point functions of the chiral primaries with symmetry $SU(2)\times SU(2) \times U(1)$ given 
in eqn.~(\ref{2F1}), see also~(\ref{Ys}). For these chiral primaries 
we can trivially perform the integration over $\phi_1$ and 
$\theta_1$ which gives a factor of $4\pi$. Furthermore, our brane embedding is characterized by $\cos \xi=1$ so we get
$Y_{\Delta}(\xi)=(-1)^{\Delta}{\cal N}_{\Delta}$ where we will ignore the overall phase.
Summarizing, the one-point functions of our chiral primaries are given by
\begin{eqnarray}
 \langle  Y_{\Delta}({x})\rangle&=& \lambda^{1/4} \,{\cal N}_{\Delta} \,{\cal C}_{\Delta} \int_0^\infty dz \,z \int_0^\infty d\rho \, \rho
 \left \{ \frac{1}{\Delta+2} \left[ \left( 3+Q^2(\Delta+5)\right) z\partial_z \right. \right. \nonumber \\
 &&+ z^2 \left(\partial_z^2 -2 \frac{Q}{z^2}\partial_z \partial_2-\frac{3}{z^3} Q \partial_2 +\frac{Q^2}{z^4} \partial_2^2
 \right) \\
 &&\ \left.\left.-\frac{1}{2} \Delta (\Delta-1)(1+Q^2) \right]+\frac{\Delta}{6} \right\} 
  \frac{1}{z^\Delta\left(\rho^2 + (x_2-{x})^2 +1/z^2\right)^\Delta}. \nonumber 
 \end{eqnarray} 
 We can now also integrate over $\rho$. This gives
 \begin{eqnarray}
 \langle  Y_{\Delta}({x})\rangle&=& \frac{\lambda^{1/4} \,{\cal N}_{\Delta} \,{\cal C}_{\Delta}}{2(\Delta-1)}
  \int_0^\infty dz \,z 
 \left \{ \frac{1}{\Delta+2} \left[ \left( 3+Q^2(\Delta+5)\right) z\partial_z \right. \right. \nonumber \\
 &&+ z^2 \left(\partial_z^2 -2 \frac{Q}{z^2}\partial_z \partial_2+\frac{3}{z^3} Q \partial_2 +\frac{Q^2}{z^4} \partial_2^2
 \right) \\
 && \left.\left.-\frac{1}{2} \Delta (\Delta-1)(1+Q^2) \right]+\frac{\Delta}{6} \right\}
  \frac{1}{z^\Delta\left( (x_2-{x})^2 +1/z^2\right)^{\Delta-1}}, \nonumber 
 \end{eqnarray} 
 provided $\Delta>1$.
 The remaining integral is not convergent for  $\Delta=2$. The same problem was observed  for the $AdS_5\times S^5$ 
 set-up~\cite{Nagasaki:2012re,Kristjansen:2012tn}. For $\Delta>2$, we find in the large-Q limit
 \begin{equation}
  \langle  Y_{\Delta}({x})\rangle \sim \lambda^{1/4}\,\frac{Q^{\Delta+1}}{{x}^{\Delta}}\sim
  \lambda^{-\Delta/2-1/4}\, \frac{q^{\Delta+1}}{ {x}^{\Delta}}.
 \end{equation}
 We notice that  ${x}$ dependence is as expected for a defect one-point function and the leading power of $q$ is the same as
 on the gauge theory side.  The powers of $\lambda$ on the two sides only agree in a large charge limit.
 

\section{Conclusion and Outlook \label{conclusion}}
 The D2-D4 domain wall version of ABJM theory shares many features with its D3-D5 counterpart in ${\cal N}=4$ super Yang Mills theory, most notably it provides us with novel examples of integrable boundary states within AdS/CFT and associated novel examples of exact overlap formulas.  These integrable boundary states generically take the form of matrix product states with the bond dimension encoding the jump in the rank of the gauge group across the domain wall and degenerate to valence bond states as when the bond dimension becomes equal to one.  Whereas the overlaps in the scalar sector of ABJM theory between Bethe eigenstates and matrix product states of bond dimension two  could be read off from a recent result on integrable boundary states in alternating $SU(N)$ spin chains~\cite{Gombor:2021hmj}, the relation to the valence bond states made it possible to extend the formula to the full theory by requiring covariance under fermionic duality transformations, thereby realizing the idea put forward in~\cite{Kristjansen:2021xno}. 
 
An obvious open problem is the extension of the overlap formula for matrix product states to general values of the bond dimension. For the integrable boundary states of relevance for ${\cal N}=4$ SYM  this could be done using representation theory of twisted Yangians~\cite{DeLeeuw:2019ohp}. It would likewise be interesting to extend the overlap formulas to higher perturbative orders and eventually to the non-perturbative situation as well. One could envision a bootstrap based strategy 
in line with~\cite{Gombor:2020kgu,Gombor:2020auk}  possibly with input from a direct perturbative calculation. The strategy for
setting up the program for perturbative calculations would follow that employed for domain wall versions of ${\cal N}=4$ SYM
where fuzzy spherical harmonics were used to disentangle the complicated mixing between color and flavor components
of the scalar fields introduced by the non-trivial vacuum expectations values~\cite{Buhl-Mortensen:2016jqo,GimenezGrau:2018jyp,Gimenez-Grau:2019fld}. 

Considering the domain wall set-up in more detail from the string theory side calls for the development of improved methods.
The strategy we used for the computation of overlaps works only for chiral primaries. As for the $AdS_5/CFT_4$ system a double scaling parameter naturally appears, but in the present case it does not allow for a precise comparison of the gauge and string theory results. One could hope that a version of supersymmetric localization makes it possible to compute overlaps
exactly for particular chiral primaries as it was the case in ${\cal N}=4$ SYM~\cite{Wang:2020seq,Komatsu:2020sup,Dedushenko:2020vgd}.  Interestingly, the operator algebra in the presence of the defect carries a natural integrability structure \cite{Dedushenko:2020yzd}. Furthermore, one could imagine an integrability based bootstrap procedure for overlaps formulated directly in the strong coupling language~\cite{Komatsu:2020sup}.
The D2-D4 probe brane set-up without flux does belongs to the list of integrable boundary 
conditions of~\cite{Dekel:2011ja} but it would be important to check that the model, like its D3-D5 probe brane cousin, remains integrable when the flux is introduced~\cite{Linardopoulos:2021rfq}.

Ultimately,  it would be interesting to arrive at a fully non-perturbative treatment of both the D2-D4 and the D3-D5 domain wall systems with the overlap formulas being given in terms of appropriate $g$-functions~\cite{Dorey:2004xk,Jiang:2019xdz}.  

Finally, it is possible that AdS/CFT contains even more integrable boundary states awaiting discovery.

\subsection*{Acknowledgements}

We thank Matthias Volk for comments on the draft. The figures were prepared using the \texttt{JaxoDraw} package \cite{Binosi:2003yf,Binosi:2008ig}.
The work of CK  was supported by DFF-FNU through grant number DFF-4002-00037. The work of KZ  was supported by the grant "Exact Results in Gauge and String Theories" from the Knut and Alice Wallenberg foundation and by RFBR grant 18-01-00460A. 
\appendix

\bibliographystyle{nb}

\end{document}